\documentclass[preprintnumbers,twocolumn,article,amsmath,amssymb,floatfix,9pt,prd,superscriptaddress,nofootinbib]{revtex4-1}
\usepackage[utf8,latin1]{inputenc}
\usepackage{graphicx}
\usepackage{dcolumn}
\usepackage[dvipsnames]{xcolor}
\usepackage[T1]{fontenc}

\usepackage{mathrsfs}  
\usepackage{cases}
\usepackage{bm}
\usepackage{academicons}
\usepackage{mathtools, nccmath}
\usepackage{fancyhdr}
\usepackage{tikz,xcolor}

\usepackage{tensor}
\usepackage[normalem]{ulem}
\usepackage{lipsum}
\usepackage{soul}
\usepackage{cancel}
\usepackage{stackengine,scalerel}
\usepackage{mathabx}
\usepackage{hyperref}
\usepackage{tabularx}
\hypersetup{colorlinks, linkcolor={red},citecolor={blue},urlcolor={blue}}  

\AtBeginDocument{\fontsize{8}{11}\selectfont}

\definecolor{lime}{HTML}{A6CE39}
\DeclareRobustCommand{\orcidicon}{
	\begin{tikzpicture}
	\draw[lime, fill=lime] (0,0) 
	circle [radius=0.16] 
	node[white] {{\fontfamily{qag}\selectfont \tiny ID}};
	\draw[white, fill=white] (-0.0625,0.095) 
	circle [radius=0.007];
	\end{tikzpicture}
	\hspace{-2mm}
}

\fancypagestyle{plain}{%
  \fancyhf{}
  \fancyfoot[C]{\iffloatpage{}{\thepage}}
  }
\foreach \x in {A, ..., Z}{%
	\expandafter\xdef\csname orcid\x\endcsname{\noexpand\href{https://orcid.org/\csname orcidauthor\x\endcsname}{\noexpand\orcidicon}}
}

\begin{document}

\title[Shadow signatures and near-horizon energy accumulation in coherent quantum sine-integral black holes]{Shadow signatures and near-horizon energy accumulation in coherent quantum sine-integral black holes}

\author{A. Errehymy\orcidA{}}
\email{abdelghani.errehymy@gmail.com (Corresponding author)}
\affiliation{Astrophysics Research Centre, School of Mathematics, Statistics and Computer Science, University of KwaZulu-Natal, Private Bag X54001, Durban 4000, South Africa}
\affiliation{Center for Theoretical Physics, Khazar University, 41 Mehseti Str., Baku, AZ1096, Azerbaijan}
\affiliation{Jadara Research Center, Jadara University, Irbid 21110, Jordan}

\author{Y. Khedif\orcidB{}}
\email{youssef.khedif@gmail.com}
\affiliation{Department of Physics, Faculty of Sciences A\"{i}n Chock, Laboratory of Mechanics and High Energy Physics, Hassan II University of Casablanca, P.O. Box 5366, 20100, Maarif Casablanca, Morocco}

\author{M. Daoud\orcidC{}}
\email{m$_{}$daoud@hotmail.com}
\affiliation{Department of Physics, Faculty of Sciences, Ibn Tofail University, P.O. Box 133, Kenitra 14000, Morocco}
\affiliation{Abdus Salam International Centre for Theoretical Physics, Miramare, Trieste 34151, Italy}

\author{B. Turimov\orcidD{}}
\email[]{bturimov@astrin.uz}
\affiliation{Engineering school, Central Asian University, Milliy bog Str.264, Tashkent, 111221,Uzbekistan}
\affiliation{University of Tashkent for Applied Sciences, Gavhar Str.1, Tashkent, 100149, Uzbekistan}
\affiliation{Ulugh Beg Astronomical Institute, Astronomy St. 33, Tashkent 100052, Uzbekistan}   

\author{S. Usanov\orcidE{}}
\email[]{sm.usanov@kiut.uz}
\affiliation{Kimyo International University in Tashkent, Shota Rustaveli Str. 156, Tashkent 100121, Uzbekistan}

\author{Z. Yasakov\orcidF{}}
\email[]{zikrillo87@mail.ru}
\affiliation{Samarkand State University of Architecture and Construction, Lolazor Street 70, 140147, Samarkand, Uzbekistan}

\author{F. Turaev\orcidI{}}
\email[]{farhodjon9618@mail.ru}
\affiliation{Alfraganus University, Yukori Karakamish Str. 2a, Tashkent 100190, Uzbekistan}

\date{\today} 

\begin{abstract}
{\footnotesize { { Quantum modifications of black hole spacetimes can lead to observable departures from the predictions of classical general relativity, particularly in the strong-gravity regime near the event horizon. In this context, the coherent quantum black hole model introduces a sine-integral correction to the gravitational potential, characterized by a coherence scale \(R_s\), which softens the central singularity and alters the near-horizon geometry. Motivated by the growing precision of black hole imaging and strong-field observations, we explore how these quantum-induced effects influence photon motion and the resulting shadow structure. Considering representative values within the interval \(R_s \in [0.5M, 1.2M]\), we show that the photon sphere shifts inward while the critical impact parameter decreases, reaching values such as \(b_{\rm crit}\approx 4.764M\) for \(R_s=0.5M\). Consequently, the shadow radius becomes systematically smaller and develops a more intricate structure than in the Schwarzschild case (\(b_{\rm crit}=3\sqrt{3}M \approx 5.196M\)). The analysis of the corresponding intensity distributions also reveals a pronounced enhancement of radiation near the shadow boundary, especially for intermediate coherence scales. This feature can be attributed to photons remaining longer within the modified near-horizon region before escaping toward distant observers. Such an increase in brightness may therefore provide a possible observational signature of quantum corrections at horizon scales. Furthermore, although the geometry allows radiation and energy to accumulate around an effective horizon surface, its response to perturbations differs from that expected for compact objects with stable light rings. Instead of contracting under backreaction effects, the accumulation surface shifts outward. These findings suggest that coherent quantum black holes preserve the asymptotic behavior of classical solutions at large distances while introducing potentially detectable imprints in strong-field gravitational lensing and black hole shadow observations.}}\\\\
\textbf{Keywords:} Quantum black holes; Null geodesics; Shadow formation; Horizon-scale quantum corrections.}
\end{abstract}

\maketitle

\section{Introduction:} \label{Sec:1}

In recent years, black hole physics has been revitalized by a series of striking observational achievements. The detection of gravitational waves from compact binary mergers by the LIGO-Virgo-KAGRA collaboration \cite{LIGOScientific2025}, along with the first horizon-scale images of supermassive black holes by the Event Horizon Telescope (EHT) \cite{EventHorizonTelescope:2019dse, EventHorizonTelescope:2022wkp}, has brought us closer than ever to directly testing gravity in its most extreme regime. These milestones have not only reinforced the success of general relativity but have also sparked renewed interest in exploring what may lie beyond it. In particular, the region close to black hole horizons has become a natural testing ground for strong-field effects and possible deviations from the classical picture \cite{Bambi2017, Barack2018, Cardoso2019, Capozziello:2023tbo}. Within this broader effort, regular black holes---solutions that avoid the formation of singularities---have gained considerable attention. Different strategies have been proposed, ranging from extensions of well-known geometries such as those of Bardeen, Dymnikova, and Hayward \cite{Zeng2025, Ma2025, Vertogradov2025b, Konoplya2024, Vertogradov2025, Fathi2021, Gohain2024, Liang2025, Capozziello:2025ycu}, to constructions based on nonlinear electrodynamics acting as an effective matter source \cite{Culetu2014,Miao2024,Sultan2025,Junior2025,Huang2025,Contreras2025}. Other ideas, including black-bounce geometries and models incorporating quantum corrections near the horizon, further highlight the variety of approaches currently under investigation \cite{Simpson2018, Xiang2013, Brahma:2020eos, Bhandari2024, Belfaqih:2024vfk, Calza2024a, Calza2024b, Calza2025c, Calza2025d, Li-S2025a}. These developments have naturally led to detailed studies of particle dynamics and quasi-periodic oscillations in various modified gravity frameworks, with recent works investigating these phenomena in charged black holes in Kalb-Ramond gravity \cite{Jumaniyozov:2025}, static Einstein-scalar-Gauss-Bonnet black holes \cite{Murodov:2026}, and non-minimally coupled Horndeski black holes \cite{Turakhonov:2026}, demonstrating the broad applicability of these probes. Alongside these theoretical efforts, recent work has explored sequential snapping bifurcations through tunable energy landscapes \cite{Zhang:2026snap}, developed new pipelines for identifying gamma-ray blazars from the 4FGL-Xiang catalog based on multiwavelength flux distributions \cite{Xiang:2026blazar}, and proposed a generalized Doppler effect for high-accuracy frequency shift measurement \cite{Zhang:2026doppler}. In parallel, studies of non-Hermitian systems have revealed transverse drift induced by the quantum metric of exceptional rings \cite{Septembre:2025drift}, demonstrated nonreciprocal spontaneous parametric processes \cite{Li:2025nonreciprocal}, and observed nonreciprocal transverse localization of light  \cite{Liang:2025localization}, collectively illustrating the rich interplay between geometry, topology, and wave propagation that also underpins many of the phenomena discussed here.

Before turning to the main analysis, it is worth noting that the observational progress mentioned above has been accompanied by a parallel effort to constrain deviations from classical black hole geometries using shadow measurements and X-ray spectroscopy. Li et al. \cite{Li:2022eue} placed constraints on the deformation of a rotating black hole mimicker from its shadow, while Mirzaev et al. \cite{Mirzaev:2023oud} simulated the shadow of a Kerr-Newman-NUT-Kiselev black hole in Rastall gravity with a thin accretion disk. More recent work by the same group has modeled X-ray spectra of black hole X-ray binaries including returning radiation \cite{Mirzaev:2024qcu} and developed a numerical parametrization of stationary axisymmetric black holes in a theory-agnostic framework \cite{Mukazhanov:2024rka}. Shashank et al. \cite{Shashank:2024hsi} extended this line of inquiry to reflection spectra of super-Eddington X-ray sources, while Errehymy and collaborators have derived constraints on quantum Oppenheimer-Snyder black holes from quasi-periodic oscillations and the S2 star orbit \cite{Errehymy:2026bbl}, on Einasto dark matter parameters from EHT images of Sgr A$^*$ and M87$^*$ \cite{Errehymy:2026wti}, and on Kazakov-Solodukhin quantum deformed black holes from the same shadow observations \cite{Errehymy:2026fgm}. These studies collectively demonstrate how increasingly precise observations are beginning to probe the spacetime geometry near the horizon and to test regular and quantum-corrected black hole models against data.

Black holes are also central to our understanding of how gravity connects with quantum physics and thermodynamics. The fact that they radiate, as predicted by Hawking, links geometry to temperature and entropy, leading to the well-known Bekenstein-Hawking relation. Within an effective field theory perspective, it becomes possible to systematically include quantum corrections, which modify both the entropy and the emission spectrum \cite{Calmet:2019eof, Ellwanger:2021mru}. From this viewpoint, the Einstein-Hilbert action is best regarded as the leading contribution in a low-energy expansion, with additional terms involving higher-order curvature invariants expected to appear at high energies. Although these corrections form an infinite series in principle, practical calculations rely on suitable truncations that still capture the dominant physical effects \cite{Odintsov:1991nd, Barvinsky:1994hw, Bonora:2014dfa, Kuntz:2019lzq, Casadio:2021rwj, Asorey:2024mkb}. Such higher-curvature contributions have been widely used to model quantum gravitational effects, influencing black hole solutions and even giving rise to potentially observable signatures, with related ideas extending to other areas like hydrodynamics through gauge/gravity duality \cite{Nastase:2016sji,Casadio:2017cdv,Casadio:2017sze,Donoghue:2018izj,Kuntz:2019omq,Wang:2019rcf,Wang:2025fmz,Calmet:2024pve,Perrucci:2024qrr,daRocha:2009gb, daRocha:2020gee, daRocha:2021xwq}.

The origin of these corrections can be traced back to early studies of quantum gravity. The one-loop analysis by 't Hooft and Veltman \cite{tHooft:1974toh} already revealed important features of quantum contributions to the Einstein-Hilbert action, while later work by Goroff and Sagnotti \cite{Goroff:1985th} showed that higher-loop terms introduce cubic curvature interactions. These developments naturally led to the construction of modified theories, such as Einstein cubic gravity and related effective models \cite{Bueno:2016lrh, Rachwal:2021bgb, DeFelice:2023vmj, Lessa:2023thi}, as well as more general frameworks that explicitly incorporate higher-order curvature terms \cite{Matyjasek:2020bzc, Calmet:2023gbw, Lambiase:2024vkz, Marciu:2023hdb}. What makes these contributions particularly relevant is that they become dominant precisely in regimes where spacetime curvature is large, offering a way to probe physics beyond the classical limit. This issue becomes unavoidable when one considers gravity at trans-Planckian scales. While general relativity remains extremely accurate in weak and moderate fields, it inevitably predicts singularities where the curvature diverges \cite{Casadio:2021gdf}. Black holes provide the clearest example of this breakdown, pointing to the need for a more complete, quantum description of spacetime. One appealing way to think about this is offered by the corpuscular, or coherent-state, picture of gravity. In this framework, spacetime is no longer viewed as fundamental but instead emerges from the collective dynamics of a large number of weakly interacting quanta \cite{Muck:2016stv, Casadio:2017cdv, Casadio:2021eio}. Black holes, in particular, can be understood as bound states of soft gravitons with occupation number \( \mathcal{N} \sim M^2  M_p^{-2} \), while the familiar classical geometry appears as an effective, averaged description of this many-body system \cite{Casadio:2015bna, Giusti:2019wdx}. Quantum effects then show up as \(\mathcal{N}^{-1}\) corrections, becoming significant exactly where the classical theory begins to fail \cite{Casadio:2016zpl,Dvali:2011aa,Dvali:2012rt}.

One important consequence of this framework is that spacetime cannot be confined to arbitrarily small regions. The finite spread of the graviton wavefunction effectively introduces a minimal length, smoothing out what would otherwise be divergent curvature. This idea fits naturally with results from effective quantum geometry, where sharp horizons are replaced by smeared structures and singularities are avoided \cite{Casadio:2015jha, Casadio:2022ndh, Cadoni:2018dnd, Casadio:2024whh, Urmanov:2024qai, Cirilo-Lombardo:2023cxb}. Further developments---including studies of quantum hair, modified collapse, and coherent quantum black hole solutions---reinforce this emerging picture \cite{Feng:2024bsx, Calmet:2023met, Casadio:2024lzd, Casadio:2023pmh, Meert:2025cct, Antonelli:2025mcv, Feng:2025nai}. Altogether, these results suggest that quantum effects can remain relevant even near the horizon, offering a consistent framework where classical behavior is recovered at large distances while singularities are naturally resolved.

A key motivation for studying regular spacetimes lies not only in their ability to smooth out curvature singularities, but also in the diverse physical effects they can exhibit. With this perspective in mind, the present work focuses, first, on the behavior of photon trajectories in quantum-corrected black holes, specifically within the framework of a coherent quantum black hole geometry written in Schwarzschild coordinates. The analysis of null geodesics provides a direct way to track how light propagates in such a background. In parallel, we investigate the corresponding shadow patterns produced by this geometry, aiming to uncover observable imprints that could distinguish it from classical scenarios. An interesting aspect of this setup is that both massive particles and light rays require an infinite affine parameter to approach the event horizon, and we therefore examine how this unusual feature manifests itself in the resulting images. Another point of interest concerns the concentration of photons and energy near \( r = 2M \), i.e., at the event horizon. This naturally raises the question of whether such an accumulation could trigger backreaction effects leading to instabilities in the spacetime, in a way reminiscent of what occurs for stable light rings in certain horizonless compact objects \cite{Keir:2014oka, Cardoso:2014sna, Cunha:2022gde}. We address this issue by emphasizing a clear distinction between those configurations and the event horizon in coherent quantum black hole geometries, showing that the underlying mechanisms differ in an essential way.

The paper is organized as follows. Sec.~\ref{Sec:II} introduces photon dynamics in coherent quantum black holes, beginning with a presentation of the coherent quantum black hole geometry in Schwarzschild coordinates in Subsec.~\ref{Sec:geometry}, followed by an analysis of photon motion in the equatorial plane in Subsec.~\ref{Sec:geodesic}, which sets the stage for studying the corresponding photon geodesics. In Sec.~\ref{Sec:III}, we examine the resulting shadow images and highlight potential observational signatures. Sec.~\ref{Sec:IV} addresses the effects of energy accumulation near the event horizon on the spacetime structure, with Subsec.~\ref{Sec:effective_geom} presenting the effective geometry with energy accumulation, Subsec.~\ref{Sec:perturbative} analyzing perturbative backreaction, Subsec.~\ref{Sec:role_qn} discussing the role of the coherent quantum potential, and Subsec.~\ref{Sec:comparison_light_rings} comparing with stable light rings. Finally, Sec.~\ref{Sec:V} summarizes our main findings.

Throughout this work, we adopt natural units with $G=c=1$.

\section{Photon Dynamics in Coherent Quantum Black Holes}\label{Sec:II}

We begin by presenting the coherent quantum black hole geometry in Schwarzschild coordinates (Subsec.~\ref{Sec:geometry}). Next, we explore photon motion in the equatorial plane (Subsec.~\ref{Sec:geodesic}), laying the groundwork for analyzing the resulting photon geodesics.

\subsection{Coherent Quantum Black Hole Geometry in Schwarzschild Coordinates}\label{Sec:geometry}

We consider a static and spherically symmetric coherent quantum black hole described in Schwarzschild coordinates by
\begin{equation}
ds^2 = -\left(1 + 2\psi_{\rm QN}\right) dt^2 + \frac{dr^2}{1 + 2\psi_{\rm QN}} + r^2 d\Omega^2,
\label{metric}
\end{equation}
where \( d\Omega^2 \) corresponds to the standard line element on the unit sphere. {The function
\begin{equation}
\psi_{\rm QN} = \frac{2 \psi_{\rm N}}{\pi} {\rm Si}\left(\frac{r}{R_s}\right),\qquad \psi_{\rm N} = -\frac{G_{\rm N} M}{r}
\label{Vqn}
\end{equation}
defines a quantum-modified gravitational potential built from the usual Newtonian potential \( \psi_{\rm N} \). Here, \(R_s\) sets the scale of the quantum core, and
\begin{equation}
{\rm Si}(x) = \int_0^x \frac{\sin z}{z}dz
\end{equation}
denotes the sine integral. As usual, \(G_{\rm N}\) is Newton's constant and \(M\) represents the ADM mass of the object.}

{For the numerical analysis and plots, we set \(M=1\) for convenience, effectively working in mass units. However, all analytic expressions retain the explicit \(M\) dependence to maintain clarity and general applicability. When \(M=1\), the mass scale is normalized such that \(R_s\) and \(b_{\rm crit}\) are expressed in units of \(M\).}

The introduction of \(R_s\) can be understood as implementing an ultraviolet cutoff that removes the divergence associated with the point-like source in the classical Schwarzschild solution. In this sense, the metric should be viewed as an effective description capturing the main features expected from a quantum-corrected compact object. Instead of a singularity, the spacetime develops a finite-size core, whose extent is typically larger than the Planck scale and comparable to the Schwarzschild radius. Depending on its value, \(R_s\) may lie below, coincide with, or exceed the horizon radius, introducing an additional degree of freedom that goes beyond the standard no-hair picture.

Far from the object (\(r \gg R_s\)), the sine integral approaches \(\pi/2\), and one recovers \(\psi_{\rm QN} \simeq \psi_{\rm N}\). In this limit, the geometry reduces to the Schwarzschild solution and remains asymptotically flat. Closer to the center, however, the behavior changes significantly: the singularity is replaced by a regularized core with an integrable central structure and finite tidal forces. Such deviations open the possibility of distinguishing this object observationally, for instance through the motion of particles or light in its vicinity.

The horizon radius \(R_H\) is obtained from the condition 
\begin{equation}\label{g^{rr}}
g^{rr} = 1 + 2\psi_{\rm QN} = 0. 
\end{equation}
Due to the oscillatory nature of the sine integral, \(R_H\) does not vary monotonically with \(R_s\), but instead fluctuates around the Schwarzschild value \(R_{\rm Sch} = 2M\). {In this case, there exist specific values of \(R_s\) for which the coherent quantum black hole horizon matches the Schwarzschild one, with the largest such value around \(R_s \simeq 1.06M\). For larger values, roughly \(R_s \gtrsim 1.2M\), the core extends beyond the horizon, leading to a horizonless compact configuration.}

To gain insight into how the quantum correction reshapes the geometry, it is useful to express the curvature invariants directly in terms of the effective potential \(\psi_{\rm QN}(r)\). In this form, the Ricci scalar can be written as
\begin{eqnarray}
R(r)=-2\psi_{\rm QN}''(r)-\frac{8}{r}\psi_{\rm QN}'(r)-\frac{4}{r^2}\psi_{\rm QN}(r).
\end{eqnarray}

The quadratic Ricci invariant takes the form
\begin{eqnarray}
R_{\mu\nu}R^{\mu\nu}(r)
&=&2\big(\psi_{\rm QN}''(r)\big)^2
+\frac{8}{r^2}\big(\psi_{\rm QN}'(r)\big)^2
+\frac{8}{r^4}\big(\psi_{\rm QN}(r)\big)^2
\nonumber\\
&+&\frac{8}{r}\psi_{\rm QN}'(r)\psi_{\rm QN}''(r),
\end{eqnarray}
while the Kretschmann scalar becomes
\begin{eqnarray}
{\cal K}(r)
=4\big(\psi_{\rm QN}''(r)\big)^2
+\frac{16}{r^2}\big(\psi_{\rm QN}'(r)\big)^2
+\frac{16}{r^4}\big(\psi_{\rm QN}(r)\big)^2.
\end{eqnarray}

{For static, spherically symmetric configurations, it is often more intuitive to describe the spacetime in terms of the Misner-Sharp-Hernandez mass function \( \mu(r) \). In this framework, the metric function satisfies (see e.g.~Ref.~\cite{Faraoni:2015ula})}
\begin{eqnarray}
1+2\psi_{\rm QN}(r)=1-\frac{2\mu(r)}{r},
\end{eqnarray}
which directly leads to
\begin{eqnarray}
\mu(r)=-r\psi_{\rm QN}(r).
\end{eqnarray}

{Using the explicit expression of the quantum potential in Eq. \eqref{Vqn}, the mass profile becomes
\begin{eqnarray}
\mu(r)=-\frac{2r\psi_{\rm N}}{\pi}{\rm Si}\left(\frac{r}{R_s}\right) = \frac{2G_{\rm N}M}{\pi}{\rm Si}\left(\frac{r}{R_s}\right).
\end{eqnarray}}

\begin{figure}
\centering
\includegraphics[width=8.9cm,height=5.9cm]{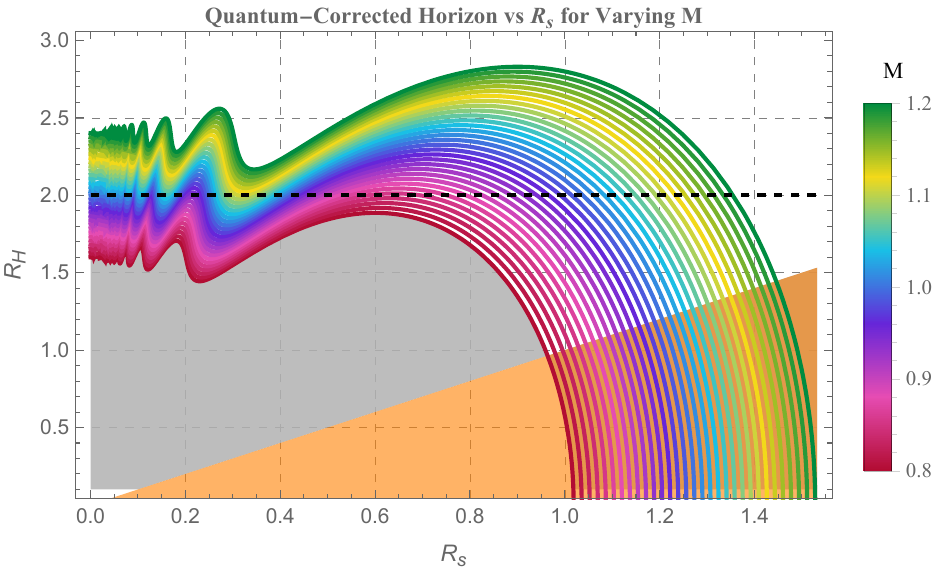}%
    \caption{\scriptsize The plot displays the horizon radius \(R_H\) as a function of the scale parameter \(R_s\) for coherent quantum black holes with a range of masses. Each colored curve corresponds to a specific mass, smoothly varying from deep red for the smallest mass to forest green for the largest, showing how the quantum-corrected horizon shifts with both \(R_s\) and mass. {The grey shaded area between the first horizon curve and the reference core (\(R = R_s\)) highlights the region where quantum effects dominate}, while the orange shading along the axis emphasizes the underlying scale reference. The thick black dashed line marks the classical Schwarzschild radius \(R_{\rm sch} = 2M\), providing a clear baseline for comparison. Grid lines on both axes help illustrate the trend and allow visual comparison of how horizon size depends on \(R_s\) and mass across the full range of coherent black holes considered.
    }    \label{Fig0}
\end{figure}

This replaces the constant Schwarzschild mass with a function that gradually increases with radius. The scale \(R_s\) controls how quickly this buildup occurs, effectively introducing a finite-sized core. {The behavior of the horizon radius \(R_H\) as a function of \(R_s\) for different black hole masses is shown in Fig.~\ref{Fig0}.}

{Close to the origin, using the small-argument expansion of the sine integral,
\begin{equation}\label{VQNEX}
{\rm Si}(x)=x-\frac{x^3}{18}+\mathcal{O}(x^5),
\end{equation}
the quantum potential behaves as
\begin{equation}
{\psi_{\rm QN}(r) = -\frac{2G_{\rm N}M}{\pi R_s} + \frac{G_{\rm N}M}{9\pi R_s^3}\,r^2 + \mathcal{O}(r^4), \qquad r\to 0^+,}
\end{equation}
{which approaches a finite constant \(\psi_{\rm QN}(0) = -2G_{\rm N}M/(\pi R_s)\) at the origin, with the leading correction being of order \(r^2\). This constant behavior of the quantum potential at \(r=0\) is consistent with the regularity of the quantum core.}

\begin{figure}
\begin{center}
\includegraphics[width=8.9cm,height=5.9cm]{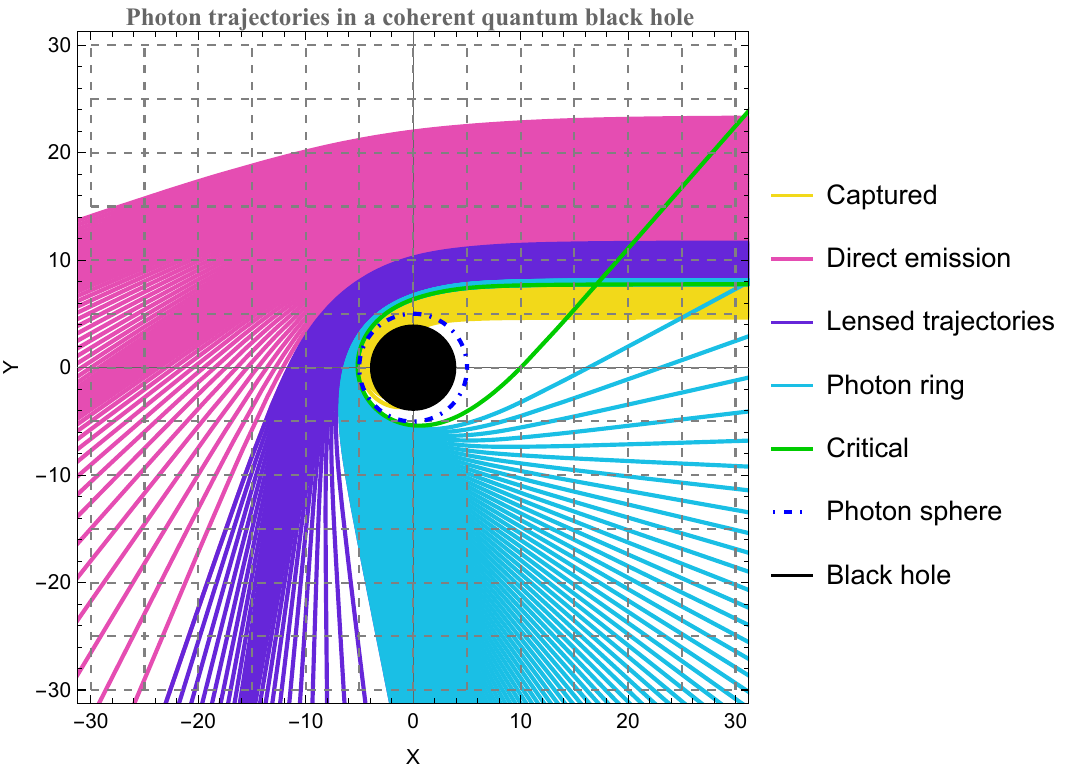}
\end{center}
\caption{\scriptsize  Light paths in the (\(X,Y\)) plane around a coherent quantum black hole with \(M=1\) and \(R_s=0.9M\). Rays are emitted from far away, and their behavior is controlled by the impact parameter \(b\) relative to the critical value \(b_c\equiv b_{crit}\). For \(b<b_c\) (yellow), photons are captured and fall into the black hole, ending inside the central dark region. The limiting case \(b=b_c\) (green) defines the boundary between capture and escape, where the trajectory winds around the photon sphere. For \(b\gtrsim b_c\) (cyan), photons loop several times near the photon sphere before escaping, forming a thin photon ring. In the range \(b_c<b\lesssim1.5\,b_c\) (violet), trajectories are strongly bent and wrap partially around the black hole. For larger values \(b\gtrsim1.5\,b_c\) (pink), the deflection becomes weak and the paths appear nearly straight. The blue dashed circle marks the photon sphere, while the black disk represents the black hole. This view clearly shows how changing \(b\) relative to \(b_c\) governs the transition from capture to escape and the formation of the photon ring.
}\label{fig:trajectory}
\end{figure}

\begin{figure}
\begin{center}
\includegraphics[width=7.9cm,height=5.9cm]{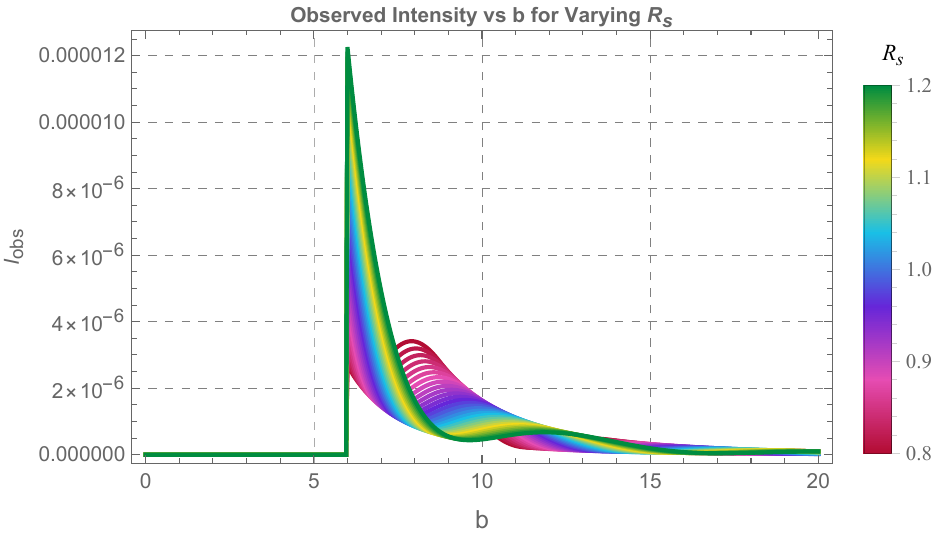}
\end{center}
\caption{\scriptsize  Observed intensity \(I_{\rm obs}\) versus impact parameter \(b\) for different source scale radii \(R_s\). Each curve represents a specific \(R_s\) between \(0.8M\) and \(1.2M\), with colors transitioning smoothly from deep red for the smallest \(R_s\) to forest green for the largest. {Yellow corresponds to the intermediate value \(R_s = 0.8M\), as indicated in the color bar.} The intensity includes the metric-inspired modulation \(G(r,R_s)^4\) and the radial emissivity profile \(I_e(r,R_s)\). Grid lines and the color bar help illustrate how the intensity changes across the \(R_s\) range.
}\label{fig:intensity}
\end{figure}

\begin{figure*}
\begin{center}
\includegraphics[width=5.9cm,height=4.9cm]{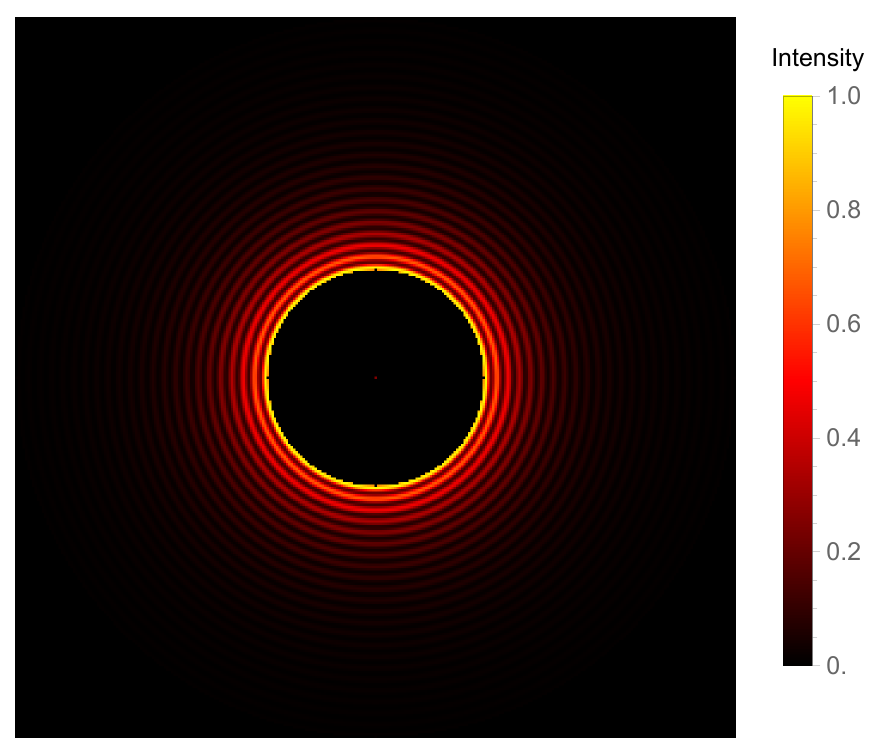}
\includegraphics[width=5.9cm,height=4.9cm]{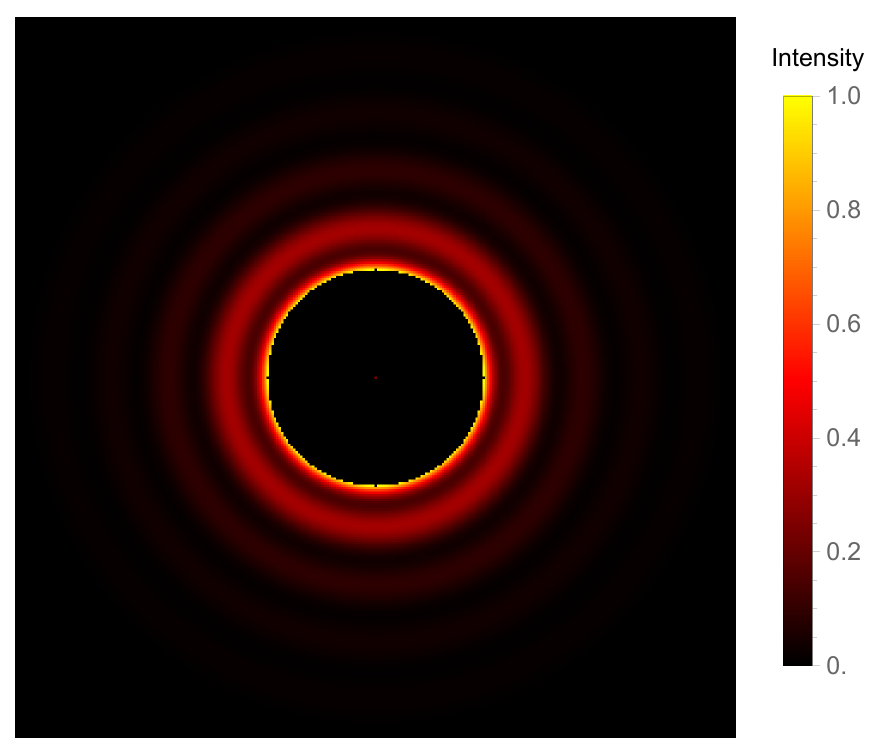}
\includegraphics[width=5.9cm,height=4.9cm]{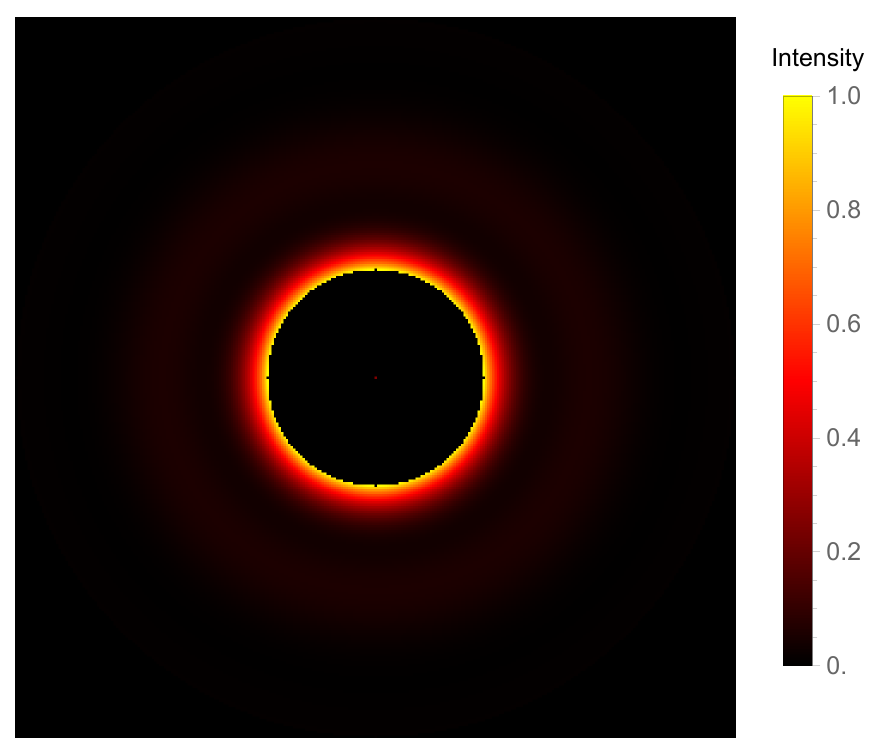}
\end{center}
\caption{\scriptsize  Intensity images of the black hole shadow cast in the spherical accretion model for three example source radii \(R_s = 0.1M\), \(0.5M\), and \(1.2M\). Each map displays the observed intensity {\(I_{\rm obs}(X,Y)\)}, calculated using the {exact redshift factor \(G(r,R_s)\) and the radial emissivity profile \(I_e(r,R_s)\)}. Colors follow a smooth gradient from deep red (low intensity) to forest green (high intensity), with the color bar indicating the intensity scale. These images illustrate how the apparent brightness distribution depends on the source scale radius, showing the transition from compact to more extended emission.
}\label{fig:shadow}
\end{figure*}

{The mass function, defined by \(\mu(r) = -r\psi_{\rm QN}(r)\), therefore behaves as}
\begin{equation}
{\mu(r) = \frac{2G_{\rm N}M}{\pi R_s}\,r + \mathcal{O}(r^3), \qquad r\to 0^+,}
\end{equation}
{so the mass smoothly vanishes at the center with a linear growth in r. This is a direct consequence of the finite, nonvanishing value of \(\psi_{\rm QN}\) at \(r=0\).}}

{Substituting these expansions into the curvature invariants gives the leading contributions}
{
\begin{eqnarray}
R(r) &=& -\frac{4\psi_{\rm QN}(0)}{r^2} + \mathcal{O}(1) = \frac{8G_{\rm N}M}{\pi R_s}\frac{1}{r^2} + \mathcal{O}(1), \\
R_{\mu\nu}R^{\mu\nu} &=& \frac{8\psi_{\rm QN}(0)^2}{r^4} + \mathcal{O}(r^{-2}) = \frac{32(G_{\rm N}M)^2}{\pi^2 R_s^2}\frac{1}{r^4} + \mathcal{O}(r^{-2}), \\
{\cal K} &=& \frac{16\psi_{\rm QN}(0)^2}{r^4} + \mathcal{O}(r^{-2}) = \frac{64(G_{\rm N}M)^2}{\pi^2 R_s^2}\frac{1}{r^4} + \mathcal{O}(r^{-2}).\qquad
\end{eqnarray}}
What emerges is a geometry with a much softer core than the classical Schwarzschild case. The mass no longer appears as a point source but instead grows smoothly from zero, reflecting an extended inner structure. Although curvature invariants still diverge as \(r\to 0\), the divergence is clearly weakened compared to the Schwarzschild case, indicating that the quantum correction reshapes the central region rather than leaving it unchanged. The effective energy density is integrable over any small spherical region, so the central singularity is mild and integrable.

The geometry's effective energy density and pressures can be written directly in terms of the quantum potential \(\psi_{\rm QN}(r)\) as
\begin{equation}
\varrho = -P_r = -\frac{\psi_{\rm QN}}{4\pi r^2} - \frac{\psi_{\rm QN}'}{4\pi r}, \quad
P_\theta = P_\phi = -\frac{\psi_{\rm QN}'}{4\pi r} - \frac{\psi_{\rm QN}''}{8\pi}.
\end{equation}
{Close to the origin, the radial energy density diverges as \(1/r^2\), while the tangential pressures remain finite.} {This divergence is milder than in the classical Schwarzschild case, and integrals over any small spherical region remain finite, so the central singularity is integrable.} This behavior is consistent with previous studies of coherent quantum black holes~\cite{Casadio:2021eio, Casadio:2022ndh}, and replacing a sharp cutoff with a smooth Gaussian regulator does not change the essential near-origin structure.

Horizons are determined by the condition \(1 + 2 \psi_{\rm QN}(R_{\rm H}) = 0\), which can also be written as
\begin{equation}
R_{\rm H} = 2 \mu(R_{\rm H}) = -2 R_{\rm H} \psi_{\rm QN}(R_{\rm H}).
\end{equation}
An event horizon exists only if the quantum scale satisfies \(R_s < 2 R_{\rm Sch}/\sqrt{\pi}\), and requiring the quantum core to lie inside the horizon imposes the stronger bound \(R_s < \mathrm{erf}(1) R_{\rm Sch}\). As \(R_s\) approaches the Schwarzschild radius, the object no longer behaves like a standard black hole, instead resembling a compact, horizonless configuration that mimics a black hole, akin to {gravastar-like} models~\cite{Mazur:2004fk}. {Fig.~\ref{Fig0} illustrates how the horizon radius \(R_H\) depends on the coherence scale \(R_s\) and black hole mass}, showing that heavier black holes have larger horizons while increasing \(R_s\) smooths the gravitational potential, reduces the horizon gradient, and highlights the region where quantum effects dominate relative to the classical Schwarzschild radius \(R_{\rm sch} = 2M\).

{Close to the origin, using the small-argument expansion of the sine integral (\ref{VQNEX}), the quantum potential tends to a finite constant, \(\psi_{\rm QN}(0) = -2G_{\rm N}M/(\pi R_s)\), while the mass function \(\mu(r) = -r\psi_{\rm QN}(r)\) grows linearly as \(\mu(r) \simeq 2G_{\rm N}M\,r/(\pi R_s)\). The radial energy density diverges as \(1/r^2\), while tangential pressures remain finite, confirming the integrability of the central singularity.} This approach naturally links the quantum potential, mass function, and spacetime curvature, offering a clear and physically consistent description of coherent quantum black holes and their horizon structure. In the following, we study Hawking radiation and the resulting energy emission, the paths of photons and their effective potential in the sine-integral coherent quantum geometry, and the stability of photon orbits as measured by the Lyapunov exponent. We also consider the weak deflection angle and test the predictions of the model against EHT observations, highlighting how all these phenomena vary with the parameter \(R_s\).

\subsection{Photon geodesics}\label{Sec:geodesic}

{ We now turn to the motion of light around a coherent quantum black hole and the resulting shadow. Since the shadow is formed entirely by photons, we restrict the analysis to massless particles. All equations in this section follow from the null condition
$g_{\mu\nu}\dot{x}^\mu\dot{x}^\nu=0$, and no massive-particle geodesics
are needed. Because the spacetime is spherically symmetric, it is enough to study light propagating in the equatorial plane, $\theta=\pi/2$. Radial motion alone does not capture the full shadow, so we consider general photon trajectories in this plane. Imposing the null condition gives
\begin{equation}
\left(1+2\psi_{\rm QN}\right)\dot{t}^2
-\frac{1}{\left(1+2\psi_{\rm QN}\right)}\dot{r}^2
-r^2\dot{\phi}^2=0,
\label{photonconstraint}
\end{equation}
where $\dot{x}^\mu=dx^\mu/d\tau$ is the tangent to the photon geodesic
$\{x^\mu(\tau)\}$ and $\tau$ is an affine parameter along the null
curve.

For photons, two quantities are conserved along the geodesic: the energy $E=-p_t$ and the angular momentum $L=p_\phi$. In terms of the photon four-momentum, these give
\begin{equation}
\frac{dt}{d\tau}=\frac{E}{1+2\psi_{\rm QN}},
\qquad
\frac{d\phi}{d\tau}=\frac{L}{r^2}.
\end{equation}

Substituting these into the null condition yields the radial equation
for photons,
\begin{equation}
\dot{r}^2
=
E^2-\frac{L^2}{r^2}\left(1+2\psi_{\rm QN}\right).
\end{equation}

It is convenient to introduce the impact parameter $b:=L/E$, which
measures how closely the light ray approaches the black hole. With this
definition, the angular motion becomes
\begin{equation}
d\phi
=
\pm
\frac{dr}{r\sqrt{\displaystyle
\frac{r^2}{b^2}-\left(1+2\psi_{\rm QN}\right)}}.
\end{equation}

The total angle swept out by a light ray passing near the black hole is obtained by integrating from infinity to the point of closest approach, $r_1$, and back. By symmetry, it is enough to consider the outward half of the trajectory, so that
\begin{equation}
\Delta\phi
=
2\int_{r_1}^{\infty}
\frac{b\,dr}{r\sqrt{r^2-b^2\left(1+2\psi_{\rm QN}\right)}},
\end{equation}
where $r_1$ is the turning point defined by $\dot{r}=0$. Changing
variables to $\epsilon=b/r$ recasts this as
\begin{equation}
\Delta\phi
=
2\int_0^{\epsilon_1}
\frac{d\epsilon}{\sqrt{
1-\epsilon^2\left(
1-\frac{4M\epsilon}{\pi b}\,
\mathrm{Si}\left(\frac{b}{\epsilon R_s}\right)
\right)
}},
\end{equation}
with $\epsilon_1=b/r_1$ and $G_{\rm N}=1$. Because of the sine integral
$\mathrm{Si}$, the integral cannot be evaluated exactly. We therefore
work in the weak-field limit, where the impact parameter is much larger
than the black hole mass, $M/b\ll1$. In this regime the sine integral
can be expanded for large arguments as
\begin{equation}
\mathrm{Si}(x)
\approx
\frac{\pi}{2}
-\frac{\cos x}{x}
-\frac{\sin x}{x^2}
+\cdots
\end{equation}
Keeping only the leading correction, the angular deviation becomes
\begin{equation}
\Delta\phi
=
2\int_0^{\epsilon_1}
\frac{d\epsilon}{\sqrt{
1-\epsilon^2\left(
1-\frac{2M\epsilon}{b}+g(\epsilon)
\right)
}},
\end{equation}
where
\begin{equation}
g(\epsilon)
=
\frac{4MR_s}{\pi b^2}\epsilon^2
\cos\left(\frac{b}{\epsilon R_s}\right).
\end{equation}
In the weak-field limit, $g(\epsilon)$ is negligible, and we recover the familiar Schwarzschild result $\Delta\phi=\pi+4M/b$. Treating
$g(\epsilon)$ as a small correction and expanding the integrand, we find
\begin{equation}
\Delta\phi
=
\pi+\frac{4M}{b}
-\frac{4MR_s}{\pi b^2}\mathcal{I}(\epsilon_1),
\end{equation}
with
\begin{equation}
\mathcal{I}(\epsilon_1)
=
\int_0^{\epsilon_1}
\frac{\epsilon^4\cos\left(\frac{b}{\epsilon R_s}\right)}
{(1-\epsilon^2)^{3/2}}
\,d\epsilon.
\end{equation}
The deflection angle is then simply $\delta\phi=\Delta\phi-\pi$.

A key quantity for the shadow is the critical impact parameter $b_{\rm crit}$, which separates photons that escape to infinity from those that are captured into the photon ring. It is determined by
\begin{equation}
\frac{1}{b_{\rm crit}^2}
-
\frac{1+2\psi_{\rm QN}(R_{\rm ph})}{R_{\rm ph}^2}
=
0,
\label{bcrit}
\end{equation}
and it directly sets the apparent size of the shadow on the observer's
sky. For a Schwarzschild black hole,
$b_{\rm crit}=3\sqrt{3}M\approx5.196M$. For a coherent quantum black
hole with $R_s=0.5M$, numerical evaluation gives
$b_{\rm crit}\approx4.764M$, a slightly smaller value.

The photon orbits can be visualized in a plane using the standard coordinates
\begin{equation}
X=r(\phi)\cos\phi,
\qquad
Y=r(\phi)\sin\phi,
\end{equation}
where $r(\phi)$ follows from the radial equation
\begin{equation}
\frac{dr}{d\phi}
=
\pm r\sqrt{
\frac{r^2}{b^2}
-\left(1+2\psi_{\rm QN}(r)\right)
}.
\end{equation}

{The resulting photon trajectories are shown in Fig.~\ref{fig:trajectory}.} The photon sphere, shown as a dashed circle in Fig.~\ref{fig:trajectory}, is the radius at which light can orbit the
black hole in an unstable circular trajectory. It is found from
\begin{equation}
\partial_r\left[
\left(1+2\psi_{\rm QN}(r)\right)r^{-2}
\right]=0.
\end{equation}
This generalizes the usual Schwarzschild condition through the quantum
correction $\psi_{\rm QN}(r)$. Far from the black hole, $\psi_{\rm QN}$
vanishes rapidly, so the photon sphere radius can be approximated as
\begin{equation}
R_{\rm PH}
\simeq
3M\left[
1-\frac{1}{2}\psi_{\rm QN}'(3M)
\right],
\end{equation}
with $\psi_{\rm QN}'>0$ in our model. As a result, the photon sphere
lies slightly closer to the black hole than in the classical case.
Similarly, expanding $b_{\rm crit}$ from Eq.~\eqref{bcrit} to first
order gives
\begin{equation}
b_{\rm crit}
\simeq
\sqrt{27}M\left[
1+\frac{1}{2}\psi_{\rm QN}(3M)
\right],
\end{equation}
which is slightly smaller than the Schwarzschild value $\sqrt{27}M$.
This shift reflects the softening of gravity near the core due to
quantum effects. The same inward shift of the photon sphere and the
critical impact parameter appears in several nonsingular black hole
models \cite{Dymnikova:1992ux, Hayward:2005gi, Nicolini:2005vd, Simpson:2019mud}, where the central singularity is effectively smoothed out. The physical picture is that the gravitational pull is weakened near the center, causing light to orbit closer in and producing a slightly smaller shadow. The coherent quantum black hole follows this same pattern, illustrating how quantum corrections modify photon paths around singularity-free spacetimes.} As seen in Fig.~\ref{fig:trajectory}, what really matters is how the impact parameter \(b\) compares to the critical value \(b_{crit}\). When \(b<b_{crit}\), the photon gets too close and gravity wins, so it cannot escape and falls into the black hole. At \(b=b_{crit}\), the situation is finely balanced: the photon neither escapes quickly nor falls in directly, but instead winds around the photon sphere, marking the transition between the two outcomes. For \(b\gtrsim b_{crit}\), the photon lingers near this region and can circle the black hole several times before finally escaping, which is what builds up the photon ring. As \(b\) becomes larger, \(b_{crit}<b\lesssim1.5\,b_{crit}\), the bending is still noticeable but less extreme, and the photon is only partially deflected. Farther away, for \(b\gtrsim1.5\,b_{crit}\), the gravitational effect becomes weak and the path is almost straight. In this way, \(b_{crit}\) naturally separates capture, strong bending, and weak deflection, showing how the black hole acts like a gravitational lens.

\section{Shadow cast}\label{Sec:III}

The groundbreaking observations by the EHT, which have captured the striking silhouettes formed by photon trajectories around M87\(^*\) and Sgr A\(^*\), have greatly advanced the study of black hole shadows across general relativity, alternative gravity theories, and quantum-gravity-inspired frameworks (see, e.g., Refs. \cite{Chen:2020aix, Chen:2022lct, Pedrotti2024, Chen:2024ibc, Wang2025a, Pantig2024, Ali2024a, Yang2025, Chen:2025jay, Al-Badawi2025, Kala2025, Chen-Yiqian2025, Pantig2025a, Wang2025b}). Shadows of black holes serve as a unique window into the strong-field regime, providing direct insight into near-horizon geometry where potential departures from general relativity could become observable. Motivated by these advances, we focus here on how the altered causal and geometric structures of coherent quantum black holes are reflected in their shadow features. 

We simulate the shadow images of coherent quantum black holes by modeling the surrounding light source as a thin, optically transparent accretion disk. As matter slowly spirals inward toward the black hole, it heats up and emits radiation. In this approach, the disk is treated as a collection of emitters distributed radially, with the emitted intensity depending only on the distance from the black hole, $r$, which greatly simplifies the computation without losing the essential physics. Assuming the disk is optically thin means that photons pass through it without being absorbed. Every time a photon intersects the disk, it effectively picks up additional brightness. Repeated intersections give rise to higher-order photon rings, which appear as layered features on the image beyond the direct emission. Despite its simplicity, this model captures the main characteristics of black hole images, and has been shown to reflect the essential phenomenology even for more complicated emission environments \cite{Gralla:2019xty, Johnson:2019ljv}. For our calculations, we consider a face-on disk orientation, consistent with what is inferred for M87\(^*\), where the inclination between the disk axis and the observer's line of sight is small \cite{CraigWalker:2018vam}. {In the numerical setup illustrated in Fig.~\ref{fig:trajectory}, the observer is placed at $(X,Y)=(40,0)$ in units where $2M = 1$, and the disk is represented by the vertical line at $X=0$.} 

{{
We now describe how the observed intensity along each photon trajectory
is computed. The starting point is the standard relativistic radiative
transfer relation, which connects the intensity in the emitter frame,
$I_{\nu_e}$, to that in the observer frame, $I_{\nu_o}$:
\begin{equation}
\frac{I_{\nu_e}}{\nu_e^3}
=
\frac{I_{\nu_o}}{\nu_o^3},
\end{equation}
which holds in the absence of absorption.

A given photon trajectory may cross the accretion disk several times.
We label the radii at which the photon intersects the disk by $r_k$,
where the index $k$ runs over the set $\mathcal{C}$ of all such
intersections. The total observed bolometric intensity is then the sum
of the contributions from every crossing, weighted by the gravitational
redshift factor $\mathcal{G}=\nu_o/\nu_e$:
\begin{equation}
I_{\rm obs}
=
\sum_{k\in\mathcal{C}}
\mathcal{G}(r_k)^4\,I_e(r_k).
\label{Iobs}
\end{equation}
This is the quantity we compute; no further approximation is made at
this stage. The set $\mathcal{C}$ contains one element for each
intersection of the photon trajectory with the disk, so a ray that
crosses the disk once contributes a single term, while a ray that wraps
around the black hole and crosses the disk multiple times contributes
several terms. This summation is what produces the characteristic
brightness enhancement near the photon ring.

The redshift factor $\mathcal{G}(r)$ encodes the relative gravitational
shift between emission at radius $r$ and observation at radius
$r_{\rm obs}$. In the spherically symmetric spacetime considered here it
is given exactly by
\begin{equation}
\mathcal{G}(r)
=
\sqrt{
\frac{1+2\psi_{\rm QN}(r_{\rm obs})}
{1+2\psi_{\rm QN}(r)}
},
\label{Gfactor}
\end{equation}
where $r_{\rm obs}$ is the observer's radial coordinate. This expression
is exact for the metric used in this work and reduces to the usual
Schwarzschild form when $\psi_{\rm QN}\to0$.

For simplicity, we assume that the emission is monochromatic in the
emitter frame at a characteristic frequency $\nu_*$,
\begin{equation}
I_{\nu_e}
=
I_e(r)\,\delta(\nu_e-\nu_*).
\end{equation}

The radial profile of the emission is modeled using the
Gralla-Lupsasca-Marrone (GLM) function \cite{Gralla:2020srx},
\begin{equation}
I_e(r;\gamma,\mu,\sigma)
=
\frac{
\exp\left\{
-\frac{1}{2}\left[
\gamma+\operatorname{arcsinh}\left(\frac{r-\mu}{\sigma}\right)
\right]^2
\right\}
}{
\sqrt{(r-\mu)^2+\sigma^2}
},
\label{GLM}
\end{equation}
where the parameters $(\gamma,\mu,\sigma)$ control the location, width,
and shape of the emission peak. Specifically, $\mu$ sets the radial
position of the peak, $\sigma$ controls its width, and $\gamma$
determines how rapidly the emission falls off away from the peak. The
GLM profile is well suited to thin-disk models because it can reproduce
a sharply peaked, asymmetric emissivity without introducing artificial
cutoffs.

For coherent quantum black holes we choose
\begin{equation}
(\gamma,\mu,\sigma)=(-3/2,\,0,\,M/2).
\label{params}
\end{equation}
This choice is motivated as follows. In thin accretion disk models the emission is expected to peak near the innermost stable circular orbit. For the quantum-corrected spacetime the innermost stable circular is slightly shifted from its Schwarzschild value, and the emission is concentrated at small radii, close to the horizon. Setting $\mu=0$ places the peak at the origin of the GLM profile, while $\sigma=M/2$ gives it a width of order the black hole mass, so that the profile is smooth and peaked near $r\sim0.5M$ and falls off rapidly for $r\gtrsim2M$. The value $\gamma=-3/2$ suppresses emission farther out, ensuring that the near-horizon region dominates the observed brightness. This is precisely the region most relevant for the shadow structure. The chosen values are representative and are intended to illustrate the qualitative features of the shadow; a full scan over $(\gamma,\mu,\sigma)$ is left for future work. We have verified that the results are robust to moderate variations of these parameters.}

{Figs.~\ref{fig:intensity} and \ref{fig:shadow} illustrate how the source scale radius $R_s$ influences the observed intensity $I_{\rm obs}$ and the intensity images of the black hole shadow cast in the spherical accretion model. Fig.~\ref{fig:intensity} shows the observed intensity $I_{\rm obs}$ as a function of the impact parameter $b$ for different source scale radii $R_s$. Each curve corresponds to a specific value of $R_s$ between $0.8M$ and $1.2M$, with colors transitioning smoothly from yellow for the smallest $R_s$ through red to forest green for the largest, as indicated in the updated color bar. The intensity combines two physical ingredients: the metric-induced modulation $\mathcal{G}(r,R_s)^4$, which encodes the gravitational redshift and the bending of photon paths in the quantum-corrected spacetime, and the radial emissivity profile $I_e(r,R_s)$, which sets how the disk emission varies with radius. Grid lines and the color bar help illustrate how the intensity changes across the range of $R_s$. Fig.~\ref{fig:shadow} presents intensity images of the black hole shadow cast in the spherical accretion model for three example source radii, $R_s=0.1M$, $0.5M$, and $1.2M$. Each map displays the observed intensity $I_{\rm obs}(X,Y)$, calculated using the exact redshift factor $\mathcal{G}(r,R_s)$ and the radial emissivity profile $I_e(r,R_s)$. Colors follow a smooth gradient from yellow for low intensity through red to forest green for high intensity, with the updated color bar indicating the intensity scale. For small $R_s$ the emission is concentrated at small radii, so the bright region is compact and peaks close to the photon ring; as $R_s$ increases, the emission extends over a wider range of radii, and the brightness distribution becomes more diffuse and spreads outward. These images therefore illustrate how the apparent brightness distribution depends on the source scale radius, showing the transition from compact to more extended emission.}}

\section{Energy accumulation and backreaction in coherent quantum black holes}\label{Sec:IV}

In the coherent quantum black hole described by Eq.~\eqref{metric}, the quantum-modified potential \( \psi_{\rm QN}(r) \) reshapes the near-horizon geometry. This feature allows radiation, particles, and energy to linger close to the horizon, somewhat like how stable light rings trap energy in horizonless objects \cite{Cunha:2017qtt, DiFilippo:2024ddg, Chen:2024ibc}. Yet, as we will see, the way the spacetime responds to this accumulation is markedly different.

\subsection{Effective geometry with energy accumulation}\label{Sec:effective_geom}
To model energy build-up, we consider a continuum of thin shells surrounding the black hole. The resulting geometry can be expressed as
\begin{align}
ds^2 = &- \big(1 + 2 \psi_{\rm QN}(r)\big)\tilde{F}(r) dt^2+
 \frac{dr^2}{\tilde{F}(r)} \nonumber \\
  &+ r^2 \left(d\theta^2 + \sin^2\theta \, d\phi^2 \right),
  \label{shellmetric_QN}
  \end{align}
where \(\tilde{F}(r) = 1 - \frac{2 \tilde{\mu}(r)}{r}\), \(\tilde{\mu}(r)\) encodes the total energy contributed by the surrounding shells.

The surface where energy effectively collects is defined by
\begin{equation}
\tilde{C}(r) := r - 2\tilde{\mu}(r) = 0.
\label{Cfunction_QN}
\end{equation}
This surface acts as an effective horizon, sensitive to the distribution of nearby energy.

\subsection{Perturbative backreaction analysis}\label{Sec:perturbative} 

To understand how the geometry reacts to additional energy, we perturb the mass function \cite{Cunha:2025oeu}:
\begin{equation}
\tilde{\mu}(r) = \tilde{\mu}_0(r) + \delta \tilde{\mu}(r),
\end{equation}
and allow the accumulation surface to shift from \( r_{\rm AS}^0 \) to
\begin{equation}
r_{\rm AS} = r_{\rm AS}^0 + \delta r_{\rm AS}.
\end{equation}

Expanding Eq.~\eqref{Cfunction_QN} to first order gives
\begin{equation}
\delta r_{\rm AS} \approx \frac{2\, \delta \tilde{\mu}(r_{\rm AS})}{d\tilde{C}/dr \, (r_{\rm AS})}.
\end{equation}

Under reasonable assumptions---that \(\tilde{C}(\infty)>0\) and \(r_{\rm AS}\) is the outermost root---we have \(\tilde{C}'(r_{\rm AS})>0\). Hence, adding energy (\(\delta \tilde{\mu}>0\)) pushes the accumulation surface outward, rather than inward.

\subsection{Role of the coherent quantum potential}\label{Sec:role_qn} 
While the location of the accumulation surface is set by \(\tilde{C}(r)=0\), the quantum potential (cf. Eq. \eqref{Vqn}) modifies the redshift structure and smooths the near-core behavior. {Near the quantum core (\(r \lesssim R_s\)), the sine integral limits the potential to a finite constant, \(\psi_{\rm QN}(r) \to -2 G_{\rm N} M/(\pi R_s)\) as \(r \to 0\), removing the divergence that appears in classical Schwarzschild geometry.} This softening allows energy to accumulate without generating a singularity.

\subsection{Comparison with stable light rings}\label{Sec:comparison_light_rings}

Energy trapping near stable light rings in horizonless objects typically leads to contraction of the trapping region \cite{Cunha:2025oeu}. In contrast, the coherent quantum horizon behaves differently: the accumulation surface expands outward as energy builds up. This highlights a key distinction between an effective horizon in a quantum-corrected geometry and conventional stable light rings.

The quantum potential introduces two crucial effects: it regularizes the near-core geometry and changes how energy backreacts on the spacetime. While the accumulation of radiation is possible near the horizon, the response is distinct from the standard stable light ring scenario. It is important to emphasize that this analysis is perturbative. Whether ongoing energy accumulation can trigger instabilities in the full nonlinear dynamics remains an open question, deserving further study.

\section{Discussion and conclusions}\label{Sec:V}

In this study, we have explored the propagation of photons in the spacetime of coherent quantum black holes and the influence of quantum modifications on the resulting shadow features. By adopting a sine-integral modification to the Schwarzschild potential (Subsec.~\ref{Sec:geometry}), we introduced a coherence scale \(R_s\) that smooths the classical singularity and produces an extended, finite core. This modification preserves the well-tested large-radius behavior of general relativity while introducing meaningful changes in the near-core region.

The central singularity of the classical Schwarzschild solution is replaced by a softer, extended core. {Near \(r=0\), the quantum potential tends to a finite constant, \(\psi_{\rm QN}(0) = -2G_{\rm N}M/(\pi R_s)\), and the mass function grows linearly from zero, \(\mu(r) \simeq 2G_{\rm N}M\,r/(\pi R_s)\). While curvature invariants still diverge, the divergence is weakened and integrable} (Subsec.~\ref{Sec:geometry}; see also Fig.~\ref{Fig0}). This behavior is consistent with other nonsingular black hole models in the literature, such as those based on nonlinear electrodynamics \cite{Dymnikova:1992ux}, Hayward regular black holes \cite{Hayward:2005gi}, and models inspired by noncommutative geometry \cite{Nicolini:2005vd}. In these frameworks, the central region is similarly regularized, suggesting that the coherent quantum metric captures essential features of singularity resolution expected from quantum gravity.

The horizon structure responds nontrivially to the coherence scale. For moderate values of \(R_s\), the horizon radius oscillates around the classical Schwarzschild value, whereas for sufficiently large \(R_s\) the object transitions into a horizonless compact configuration (Fig.~\ref{Fig0}). This reinforces parallels with gravastar-like models \cite{Mazur:2004fk} and other horizonless solutions that mimic black holes without true event horizons. These results demonstrate that quantum modifications can produce compact, nonsingular objects that remain observationally viable.

Photon motion in this quantum-corrected spacetime departs from the Schwarzschild case in measurable ways (Subsec.~\ref{Sec:geodesic}). The photon sphere is shifted inward, and the critical impact parameter \(b_{\rm crit}\) is reduced relative to the classic value of \(3\sqrt{3}M\). As shown in Fig.~\ref{fig:trajectory}, this modification alters the classification of photon trajectories: photons with \(b<b_{\rm crit}\) are captured, those with \(b\approx b_{\rm crit}\) linger near the photon sphere before escaping, and larger \(b\) yield progressively weaker bending. This pattern closely parallels trends seen in other regular black hole solutions where quantum effects soften the gravitational potential near the core \cite{Simpson:2019mud}.

Shadow intensity maps further illustrate how the coherence scale imprints on observables. As seen in Figs.~\ref{fig:intensity} and \ref{fig:shadow}, smaller \(R_s\) produces compact, high-contrast shadow features, whereas larger \(R_s\) leads to broader, more diffuse emission patterns. These results echo findings in studies of black hole images under alternate geometries and emission models \cite{Chen:2020aix, Gralla:2019xty, Johnson:2019ljv}. Importantly, even in the weak-field limit, the sine-integral correction contributes small oscillatory deviations to the deflection angle, suggesting potential signatures in precision lensing measurements that extend beyond the classical Schwarzschild prediction.

{Compared to the previous study in Ref. [75], our work provides several new insights: (i) a more accurate determination of the critical impact parameter including explicit \(M\)-dependence, yielding \(b_{\rm crit} \approx 4.764M\) for \(R_s = 0.5M\) compared to the Schwarzschild value \(3\sqrt{3}M \approx 5.196M\); (ii) a systematic analysis of the observed intensity distribution for a range of \(R_s\) values between \(0.5M\) and \(1.2M\), revealing the dependence of the shadow morphology on the quantum coherence scale; (iii) a perturbative analysis of energy accumulation and backreaction near the horizon, showing that the accumulation surface expands outward when energy is added, in contrast to the contraction observed for stable light rings; and (iv) a clearer classification of photon trajectories based on the impact parameter relative to \(b_{\rm crit}\), distinguishing capture, strong bending, and weak deflection regimes. These results extend and refine the findings of Ref. [75] by providing a more complete picture of the observational signatures of coherent quantum black holes.}

Overall, coherent quantum black holes provide a unified framework in which quantum effects regularize the near-core singularity without compromising the familiar behavior at large distances. Photon trajectories, shadow morphology, and lensing properties are all modified in ways that are physically interpretable and, in principle, observable with horizon-scale instruments such as the EHT (Sec.~\ref{Sec:III}). This work sets the stage for future investigations into rotating versions of the model, the influence of different emission environments, and direct quantitative comparison with observational data.

In summary, the coherent quantum black hole model demonstrates that singularity-free spacetimes leave clear imprints on photon dynamics and shadows. By linking quantum corrections to observable behavior (Figs.~\ref{fig:trajectory}-\ref{fig:shadow}), our results suggest a promising avenue to test quantum gravity effects in real astrophysical systems.

\section*{Acknowledgments}
{We thank the anonymous reviewers for their valuable comments and detailed suggestions, which have greatly helped to improve the clarity and quality of this work. }\\

\section*{Conflict Of Interest statement } 
No conflict of interest declared by the authors.\\

\section*{Data Availability Statement:}  
{This manuscript has no associated data, or the data will not be deposited. (There is no observational data related to this article. The necessary calculations and graphic discussion can be made available on request.)}

\bibliography{references}

@article{Capozziello:2023tbo,
    author = "Capozziello, Salvatore and Zare, Soroush and Nieto, Luis M. and Hassanabadi, Hassan",
    title = "{Modified Kerr black holes surrounded by dark matter spike}",
    eprint = "2311.12896",
    archivePrefix = "arXiv",
    primaryClass = "gr-qc",
    doi = "10.1016/j.dark.2025.102065",
    journal = "Phys. Dark Univ.",
    volume = "50",
    pages = "102065",
    year = "2025"
}

@article{Capozziello:2025ycu,
    author = "Capozziello, Salvatore and Gambino, Serena and Luongo, Orlando",
    title = "{Comparing Bondi and Novikov{\textendash}Thorne accretion disk luminosity around regular black holes}",
    eprint = "2503.21987",
    archivePrefix = "arXiv",
    primaryClass = "gr-qc",
    doi = "10.1016/j.dark.2025.101950",
    journal = "Phys. Dark Univ.",
    volume = "48",
    pages = "101950",
    year = "2025"
}

@article{Johnson:2019ljv,
    author = "Johnson, Michael D. and others",
    title = "{Universal interferometric signatures of a black hole{\textquoteright}s photon ring}",
    eprint = "1907.04329",
    archivePrefix = "arXiv",
    primaryClass = "astro-ph.IM",
    doi = "10.1126/sciadv.aaz1310",
    journal = "Sci. Adv.",
    volume = "6",
    number = "12",
    pages = "eaaz1310",
    year = "2020"
}

@article{Cunha:2022gde,
    author = "Cunha, Pedro V. P. and Herdeiro, Carlos and Radu, Eugen and Sanchis-Gual, Nicolas",
    title = "{Exotic Compact Objects and the Fate of the Light-Ring Instability}",
    eprint = "2207.13713",
    archivePrefix = "arXiv",
    primaryClass = "gr-qc",
    doi = "10.1103/PhysRevLett.130.061401",
    journal = "Phys. Rev. Lett.",
    volume = "130",
    number = "6",
    pages = "061401",
    year = "2023"
}

@article{DiFilippo:2024ddg,
    author = "Di Filippo, Francesco",
    title = "{Nature of inner light rings}",
    eprint = "2404.07357",
    archivePrefix = "arXiv",
    primaryClass = "gr-qc",
    reportNumber = "Phys. Rev. D 110, 084026",
    doi = "10.1103/PhysRevD.110.084026",
    journal = "Phys. Rev. D",
    volume = "110",
    number = "8",
    pages = "084026",
    year = "2024"
}

@article{Cardoso:2014sna,
    author = "Cardoso, Vitor and Crispino, Lu{\'\i}s C. B. and Macedo, Caio F. B. and Okawa, Hirotada and Pani, Paolo",
    title = "{Light rings as observational evidence for event horizons: long-lived modes, ergoregions and nonlinear instabilities of ultracompact objects}",
    eprint = "1406.5510",
    archivePrefix = "arXiv",
    primaryClass = "gr-qc",
    doi = "10.1103/PhysRevD.90.044069",
    journal = "Phys. Rev. D",
    volume = "90",
    number = "4",
    pages = "044069",
    year = "2014"
}

@article{Keir:2014oka,
    author = "Keir, Joe",
    title = "{Slowly decaying waves on spherically symmetric spacetimes and ultracompact neutron stars}",
    eprint = "1404.7036",
    archivePrefix = "arXiv",
    primaryClass = "gr-qc",
    doi = "10.1088/0264-9381/33/13/135009",
    journal = "Class. Quant. Grav.",
    volume = "33",
    number = "13",
    pages = "135009",
    year = "2016"
}

@article{Cunha:2017qtt,
    author = "Cunha, Pedro V. P. and Berti, Emanuele and Herdeiro, Carlos A. R.",
    title = "{Light-Ring Stability for Ultracompact Objects}",
    eprint = "1708.04211",
    archivePrefix = "arXiv",
    primaryClass = "gr-qc",
    doi = "10.1103/PhysRevLett.119.251102",
    journal = "Phys. Rev. Lett.",
    volume = "119",
    number = "25",
    pages = "251102",
    year = "2017"
}

@article{Belfaqih:2024vfk,
    author = "Belfaqih, Idrus Husin and Bojowald, Martin and Brahma, Suddhasattwa and Duque, Erick I.",
    title = "{Black holes in effective loop quantum gravity: Covariant holonomy modifications}",
    eprint = "2407.12087",
    archivePrefix = "arXiv",
    primaryClass = "gr-qc",
    doi = "10.1103/1tyh-87sr",
    journal = "Phys. Rev. D",
    volume = "112",
    number = "4",
    pages = "046022",
    year = "2025"
}

@article{Chen:2020aix,
    author = "Chen, Che-Yu",
    title = "{Rotating black holes without $\mathbb{Z}_2$ symmetry and their shadow images}",
    eprint = "2004.01440",
    archivePrefix = "arXiv",
    primaryClass = "gr-qc",
    doi = "10.1088/1475-7516/2020/05/040",
    journal = "JCAP",
    volume = "05",
    pages = "040",
    year = "2020"
}

@article{Chen:2022lct,
    author = "Chen, Che-Yu",
    title = "{Testing black hole equatorial reflection symmetry using Sgr{\,}A* shadow images}",
    eprint = "2205.06962",
    archivePrefix = "arXiv",
    primaryClass = "gr-qc",
    doi = "10.1103/PhysRevD.106.044009",
    journal = "Phys. Rev. D",
    volume = "106",
    number = "4",
    pages = "044009",
    year = "2022"
}

@article{Brahma:2020eos,
    author = "Brahma, Suddhasattwa and Chen, Che-Yu and Yeom, Dong-han",
    title = "{Testing Loop Quantum Gravity from Observational Consequences of Nonsingular Rotating Black Holes}",
    eprint = "2012.08785",
    archivePrefix = "arXiv",
    primaryClass = "gr-qc",
    doi = "10.1103/PhysRevLett.126.181301",
    journal = "Phys. Rev. Lett.",
    volume = "126",
    number = "18",
    pages = "181301",
    year = "2021"
}

@article{Chen:2024ibc,
    author = "Chen, Che-Yu and Yokokura, Yuki",
    title = "{Imaging a semiclassical horizonless compact object with strong redshift}",
    eprint = "2403.09388",
    archivePrefix = "arXiv",
    primaryClass = "gr-qc",
    reportNumber = "RIKEN-iTHEMS-Report-24",
    doi = "10.1103/PhysRevD.109.104058",
    journal = "Phys. Rev. D",
    volume = "109",
    number = "10",
    pages = "104058",
    year = "2024"
}

@article{Chen:2025jay,
    author = "Chen, Che-Yu and Chen, Chiang-Mei and Ohta, Nobuyoshi",
    title = "{Shadow of black holes with consistent thermodynamics}",
    eprint = "2510.00708",
    archivePrefix = "arXiv",
    primaryClass = "gr-qc",
    reportNumber = "RIKEN-iTHEMS-Report-25, NITEP 263",
    month = "10",
    year = "2025"
}

@article{Nicolini:2005vd,
    author = "Nicolini, Piero and Smailagic, Anais and Spallucci, Euro",
    title = "{Noncommutative geometry inspired Schwarzschild black hole}",
    eprint = "gr-qc/0510112",
    archivePrefix = "arXiv",
    doi = "10.1016/j.physletb.2005.11.004",
    journal = "Phys. Lett. B",
    volume = "632",
    pages = "547--551",
    year = "2006"
}

@article{Hayward:2005gi,
    author = "Hayward, Sean A.",
    title = "{Formation and evaporation of regular black holes}",
    eprint = "gr-qc/0506126",
    archivePrefix = "arXiv",
    doi = "10.1103/PhysRevLett.96.031103",
    journal = "Phys. Rev. Lett.",
    volume = "96",
    pages = "031103",
    year = "2006"
}

@article{Simpson:2019mud,
    author = "Simpson, Alex and Visser, Matt",
    title = "{Regular black holes with asymptotically Minkowski cores}",
    eprint = "1911.01020",
    archivePrefix = "arXiv",
    primaryClass = "gr-qc",
    doi = "10.3390/universe6010008",
    journal = "Universe",
    volume = "6",
    number = "1",
    pages = "8",
    year = "2019"
}

@article{Dymnikova:1992ux,
    author = "Dymnikova, I.",
    title = "{Vacuum nonsingular black hole}",
    doi = "10.1007/BF00760226",
    journal = "Gen. Rel. Grav.",
    volume = "24",
    pages = "235--242",
    year = "1992"
}

@article{Cunha:2025oeu,
    author = "Cunha, Pedro V. P.",
    title = "{Backreaction of perturbations around a stable light ring}",
    eprint = "2503.00117",
    archivePrefix = "arXiv",
    primaryClass = "gr-qc",
    doi = "10.1088/1475-7516/2025/05/083",
    journal = "JCAP",
    volume = "05",
    pages = "083",
    year = "2025"
}

@article{CraigWalker:2018vam,
    author = "Craig Walker, R. and Hardee, Phillip E. and Davies, Frederick B. and Ly, Chun and Junor, William",
    title = "{The Structure and Dynamics of the Subparsec Jet in M87 Based on 50 VLBA Observations over 17 Years at 43 GHz}",
    eprint = "1802.06166",
    archivePrefix = "arXiv",
    primaryClass = "astro-ph.HE",
    doi = "10.3847/1538-4357/aaafcc",
    journal = "Astrophys. J.",
    volume = "855",
    number = "2",
    pages = "128",
    year = "2018"
}

@article{Gralla:2020srx,
    author = "Gralla, Samuel E. and Lupsasca, Alexandru and Marrone, Daniel P.",
    title = "{The shape of the black hole photon ring: A precise test of strong-field general relativity}",
    eprint = "2008.03879",
    archivePrefix = "arXiv",
    primaryClass = "gr-qc",
    doi = "10.1103/PhysRevD.102.124004",
    journal = "Phys. Rev. D",
    volume = "102",
    number = "12",
    pages = "124004",
    year = "2020"
}

@article{Gralla:2019xty,
    author = "Gralla, Samuel E. and Holz, Daniel E. and Wald, Robert M.",
    title = "{Black Hole Shadows, Photon Rings, and Lensing Rings}",
    eprint = "1906.00873",
    archivePrefix = "arXiv",
    primaryClass = "astro-ph.HE",
    doi = "10.1103/PhysRevD.100.024018",
    journal = "Phys. Rev. D",
    volume = "100",
    number = "2",
    pages = "024018",
    year = "2019"
}

@article{Pedrotti2024,
    author = "Pedrotti, Davide and Vagnozzi, Sunny",
    title = "{Quasinormal modes-shadow correspondence for rotating regular black holes}",
    eprint = "2404.07589",
    archivePrefix = "arXiv",
    primaryClass = "gr-qc",
    doi = "10.1103/PhysRevD.110.084075",
    journal = "Phys. Rev. D",
    volume = "110",
    number = "8",
    pages = "084075",
    year = "2024"
}

@book{Bambi2017,
    author = "Bambi, Cosimo",
    title = "{Black Holes: A Laboratory for Testing Strong Gravity}",
    doi = "10.1007/978-981-10-4524-0",
    isbn = "978-981-10-4523-3, 978-981-13-5158-7, 978-981-10-4524-0",
    publisher = "Springer",
    year = "2017"
}

@article{Liang2025,
    author = "Liang, Qi-Qi and Cai, Ziqiang and Liu, Dong and Long, Zheng-Wen",
    title = "{Observational properties and quasinormal Modes of the Hayward black Hole surrounded by a cloud of strings}",
    eprint = "2511.02396",
    archivePrefix = "arXiv",
    primaryClass = "gr-qc",
    month = "11",
    year = "2025"
}

@article{Konoplya2024,
    author = "Konoplya, R. A. and Zhidenko, A.",
    title = "{Dymnikova black hole from an infinite tower of higher-curvature corrections}",
    eprint = "2404.09063",
    archivePrefix = "arXiv",
    primaryClass = "gr-qc",
    doi = "10.1016/j.physletb.2024.138945",
    journal = "Phys. Lett. B",
    volume = "856",
    pages = "138945",
    year = "2024"
}

@article{Ma2025,
    author = "Ma, Meng-Sen and He, Yun and Wang, Xiao-Ming and Li, Huai-Fan",
    title = "{From singular to regular: Revisiting thermodynamics of Bardeen-Ads black holes}",
    eprint = "2510.06576",
    archivePrefix = "arXiv",
    primaryClass = "hep-th",
    doi = "10.1016/j.physletb.2025.139961",
    journal = "Phys. Lett. B",
    volume = "870",
    pages = "139961",
    year = "2025"
}

@article{Culetu2014,
    author = "Culetu, Hristu",
    title = "{On a regular charged black hole with a nonlinear electric source}",
    eprint = "1408.3334",
    archivePrefix = "arXiv",
    primaryClass = "gr-qc",
    doi = "10.1007/s10773-015-2521-6",
    journal = "Int. J. Theor. Phys.",
    volume = "54",
    number = "8",
    pages = "2855--2863",
    year = "2015"
}

@article{Simpson2018,
    author = "Simpson, Alex and Visser, Matt",
    title = "{Black-bounce to traversable wormhole}",
    eprint = "1812.07114",
    archivePrefix = "arXiv",
    primaryClass = "gr-qc",
    doi = "10.1088/1475-7516/2019/02/042",
    journal = "JCAP",
    volume = "02",
    pages = "042",
    year = "2019"
}

@article{Barack2018,
    author = "Barack, Leor and others",
    title = "{Black holes, gravitational waves and fundamental physics: a roadmap}",
    eprint = "1806.05195",
    archivePrefix = "arXiv",
    primaryClass = "gr-qc",
    doi = "10.1088/1361-6382/ab0587",
    journal = "Class. Quant. Grav.",
    volume = "36",
    number = "14",
    pages = "143001",
    year = "2019"
}

@article{Cardoso2019,
    author = "Cardoso, Vitor and Pani, Paolo",
    title = "{Testing the nature of dark compact objects: a status report}",
    eprint = "1904.05363",
    archivePrefix = "arXiv",
    primaryClass = "gr-qc",
    doi = "10.1007/s41114-019-0020-4",
    journal = "Living Rev. Rel.",
    volume = "22",
    number = "1",
    pages = "4",
    year = "2019"
}

@article{Miao2024,
    author = "Miao, Wei-Jie and Yang, Si-Jiang",
    title = "{Overcharging a nonsingular black hole in general relativity: the nonlinear electrodynamic field effects}",
    eprint = "2409.07305",
    archivePrefix = "arXiv",
    primaryClass = "gr-qc",
    doi = "10.1088/1475-7516/2025/05/022",
    journal = "JCAP",
    volume = "05",
    pages = "022",
    year = "2025"
}

@article{Gohain2024,
    author = "Gohain, Mrinnoy M. and Bhuyan, Kalyan and Borgohain, Rajnandini and Gogoi, Tonmoyee and Bhuyan, Kakoli and Phukon, Prabwal",
    title = "{Frolov black hole surrounded by quintessence - I: Thermodynamics, geodesics and shadows}",
    eprint = "2412.06252",
    archivePrefix = "arXiv",
    primaryClass = "gr-qc",
    doi = "10.1016/j.nuclphysb.2025.117073",
    journal = "Nucl. Phys. B",
    volume = "1018",
    pages = "117073",
    year = "2025"
}

@article{Huang2025,
    author = "Huang, Hyat and Rao, Xiao-Ping",
    title = "{Regular black holes and their singular families}",
    eprint = "2503.13133",
    archivePrefix = "arXiv",
    primaryClass = "gr-qc",
    doi = "10.1103/PhysRevD.111.104040",
    journal = "Phys. Rev. D",
    volume = "111",
    number = "10",
    pages = "104040",
    year = "2025"
}

@article{Vertogradov2025,
    author = "Vertogradov, Vitalii",
    title = "{Gravitational collapse and formation of regular black holes: Dymnikova, Hayward, and beyond}",
    eprint = "2504.19292",
    archivePrefix = "arXiv",
    primaryClass = "gr-qc",
    doi = "10.1140/epjc/s10052-025-14583-2",
    journal = "Eur. Phys. J. C",
    volume = "85",
    number = "8",
    pages = "839",
    year = "2025"
}

@article{Junior2025,
    author = "Junior, Ednaldo L. B. and Junior, Jos{\'e} Tarciso S. S. and Lobo, Francisco S. N. and Rodrigues, Manuel E. and da Silva, Lu{\'\i}s F. Dias and Vieira, Henrique A.",
    title = "{Dyonic regular black bounce solutions in general relativity}",
    eprint = "2502.13327",
    archivePrefix = "arXiv",
    primaryClass = "gr-qc",
    doi = "10.1140/epjc/s10052-025-14427-z",
    journal = "Eur. Phys. J. C",
    volume = "85",
    number = "7",
    pages = "724",
    year = "2025"
}

@article{Sultan2025,
    author = "Sultan, Muhammad Danish and Chaudhary, Shahid and Anwar, Talha and Ashraf, Asifa and Mubaraki, Ali M. and Atamurotov, Farruh and Abidi, Awatef",
    title = "{Thermal phase transitions, evaporation, and shadows of regular black holes in extended gravity with stringy matter}",
    doi = "10.1016/j.dark.2025.102077",
    journal = "Phys. Dark Univ.",
    volume = "50",
    pages = "102077",
    year = "2025"
}

@article{Contreras2025,
    author = "Contreras, Ernesto and Carrasco-Hidalgo, Mikaela and Bargue{\~n}o, Pedro and Suvorov, Arthur G.",
    title = "{General framework for the spontaneous scalarization of regular black holes}",
    eprint = "2511.01544",
    archivePrefix = "arXiv",
    primaryClass = "gr-qc",
    doi = "10.1103/9st9-ys67",
    journal = "Phys. Rev. D",
    volume = "112",
    number = "12",
    pages = "124053",
    year = "2025"
}

@article{Calza2024a,
    author = "Calz{\`a}, Marco and Pedrotti, Davide and Vagnozzi, Sunny",
    title = "{Primordial regular black holes as all the dark matter. I. Time-radial-symmetric metrics}",
    eprint = "2409.02804",
    archivePrefix = "arXiv",
    primaryClass = "gr-qc",
    doi = "10.1103/PhysRevD.111.024009",
    journal = "Phys. Rev. D",
    volume = "111",
    number = "2",
    pages = "024009",
    year = "2025"
}

@article{Calza2024b,
    author = "Calz{\`a}, Marco and Pedrotti, Davide and Vagnozzi, Sunny",
    title = "{Primordial regular black holes as all the dark matter. II. Non-time-radial-symmetric and loop quantum gravity-inspired metrics}",
    eprint = "2409.02807",
    archivePrefix = "arXiv",
    primaryClass = "gr-qc",
    doi = "10.1103/PhysRevD.111.024010",
    journal = "Phys. Rev. D",
    volume = "111",
    number = "2",
    pages = "024010",
    year = "2025"
}

@article{Calza2025c,
    author = "Calz{\`a}, Marco and Pedrotti, Davide and Yuan, Guan-Wen and Vagnozzi, Sunny",
    title = "{Primordial regular black holes as all the dark matter. III. Covariant canonical quantum gravity models}",
    eprint = "2507.02396",
    archivePrefix = "arXiv",
    primaryClass = "gr-qc",
    doi = "10.1103/4x1f-vctx",
    month = "7",
    year = "2025"
}

@article{Yang2025,
    author = "Yang, Chen-Yu and Ye, Huan and Zeng, Xiao-Xiong",
    title = "{Shadow and Polarization Images of Rotating Black Holes in Kalb-Ramond Gravity Illuminated by Several Thick Accretion Disks}",
    eprint = "2510.21229",
    archivePrefix = "arXiv",
    primaryClass = "gr-qc",
    month = "10",
    year = "2025"
}

@article{Chen-Yiqian2025,
    author = "Chen, Yiqian and Guo, Guangzhou and Mu, Benrong and Wang, Peng",
    title = "{Schwarzschild Black Holes Immersed in Born-Infeld Magnetic Fields and Their Observational Signatures}",
    eprint = "2506.19581",
    archivePrefix = "arXiv",
    primaryClass = "gr-qc",
    month = "6",
    year = "2025"
}

@article{Xiang2013,
    author = "Xiang, Li and Ling, Yi and Shen, You Gen",
    title = "{Singularities and the Finale of Black Hole Evaporation}",
    eprint = "1305.3851",
    archivePrefix = "arXiv",
    primaryClass = "gr-qc",
    doi = "10.1142/S0218271813420169",
    journal = "Int. J. Mod. Phys. D",
    volume = "22",
    pages = "1342016",
    year = "2013"
}

@article{LIGOScientific2025,
    author = "Abac, A. G. and others",
    collaboration = "LIGO Scientific, KAGRA, VIRGO",
    title = "{GWTC-4.0: An Introduction to Version 4.0 of the Gravitational-Wave Transient Catalog}",
    eprint = "2508.18080",
    archivePrefix = "arXiv",
    primaryClass = "gr-qc",
    reportNumber = "LIGO-P2400293",
    doi = "10.3847/2041-8213/ae0c06",
    journal = "Astrophys. J. Lett.",
    volume = "995",
    number = "1",
    pages = "L18",
    year = "2025"
}

@article{Ali2024a,
    author = "Ali, Riasat and Tiecheng, Xia and Awais, Muhammad and Babar, Rimsha",
    title = "{Deflection angle and shadow evolution from charged torus-like black hole under the effect of non-magnetic plasma and non-plasma medium}",
    doi = "10.1142/S0219887824501809",
    journal = "Int. J. Geom. Meth. Mod. Phys.",
    volume = "21",
    number = "11",
    pages = "2450180",
    year = "2024"
}

@article{Zeng2025,
    author = "Zeng, Hui and Meng, Yuan",
    title = "{Images from disk and spherical accretions of Bardeen black hole surrounded by perfect fluid dark matter}",
    eprint = "2512.05147",
    archivePrefix = "arXiv",
    primaryClass = "gr-qc",
    month = "12",
    year = "2025"
}

@article{Li-S2025a,
    author = "Li, Shulan and Wu, Jian-Pin and Ge, Xian-Hui",
    title = "{Non-singular cosmologies matching regular black holes}",
    eprint = "2512.00926",
    archivePrefix = "arXiv",
    primaryClass = "gr-qc",
    month = "11",
    year = "2025"
}

@article{Vertogradov2025b,
    author = "Vertogradov, Vitalii and Rincon, Angel",
    title = "{Energy extraction and evolution of regular black holes: The case of Bardeen spacetime}",
    eprint = "2508.14489",
    archivePrefix = "arXiv",
    primaryClass = "gr-qc",
    doi = "10.1016/j.dark.2025.102066",
    journal = "Phys. Dark Univ.",
    volume = "50",
    pages = "102066",
    year = "2025"
}

@article{Pantig2024,
    author = {Pantig, Reggie C. and Kala, Shubham and {\"O}vg{\"u}n, Ali and Lobos, Nikko John Leo S.},
    title = "{Testing black holes with cosmological constant in Einstein-bumblebee gravity through the black hole shadow using EHT data and deflection angle}",
    eprint = "2410.13661",
    archivePrefix = "arXiv",
    primaryClass = "gr-qc",
    doi = "10.1142/S0219887825502408",
    month = "10",
    year = "2024"
}

@article{Pantig2025a,
    author = {Pantig, Reggie C. and {\"O}vg{\"u}n, Ali},
    title = "{Multimodal signatures of asymptotic (A)dS Kalb{\textendash}Ramond black holes: Constraints through the shadow, weak deflection angle, and topological photon spheres}",
    eprint = "2503.18585",
    archivePrefix = "arXiv",
    primaryClass = "gr-qc",
    doi = "10.1016/j.aop.2025.170104",
    journal = "Annals Phys.",
    volume = "480",
    pages = "170104",
    year = "2025"
}

@article{Kala2025,
    author = "Kala, Shubham and Singh, Jaswinder",
    title = "{Gravitational lensing and shadow around a non-minimally coupled Horndeski black hole in plasma medium}",
    eprint = "2507.17280",
    archivePrefix = "arXiv",
    primaryClass = "gr-qc",
    doi = "10.1140/epjc/s10052-025-14793-8",
    journal = "Eur. Phys. J. C",
    volume = "85",
    number = "9",
    pages = "1047",
    year = "2025"
}

@article{Al-Badawi2025,
    author = "Al-Badawi, Ahmad",
    title = "{The shadow of black holes in F(R)-ModMax theory with cosmic strings}",
    eprint = "2510.22704",
    archivePrefix = "arXiv",
    primaryClass = "gr-qc",
    doi = "10.1209/0295-5075/ae075c",
    journal = "EPL",
    volume = "152",
    number = "1",
    pages = "19002",
    year = "2025"
}

@article{Wang2025b,
    author = "Wang, Zi-Liang and Battista, Emmanuele",
    title = "{Dynamical features and shadows of quantum Schwarzschild black hole in effective field theories of gravity}",
    eprint = "2501.14516",
    archivePrefix = "arXiv",
    primaryClass = "gr-qc",
    doi = "10.1140/epjc/s10052-025-13833-7",
    journal = "Eur. Phys. J. C",
    volume = "85",
    number = "3",
    pages = "304",
    year = "2025"
}

@article{EventHorizonTelescope:2022wkp,
    author = "Akiyama, Kazunori and others",
    collaboration = "Event Horizon Telescope",
    title = "{First Sagittarius A* Event Horizon Telescope Results. I. The Shadow of the Supermassive Black Hole in the Center of the Milky Way}",
    eprint = "2311.08680",
    archivePrefix = "arXiv",
    primaryClass = "astro-ph.HE",
    doi = "10.3847/2041-8213/ac6674",
    journal = "Astrophys. J. Lett.",
    volume = "930",
    number = "2",
    pages = "L12",
    year = "2022"
}

@article{EventHorizonTelescope:2019dse,
    author = "Akiyama, Kazunori and others",
    collaboration = "Event Horizon Telescope",
    title = "{First M87 Event Horizon Telescope Results. I. The Shadow of the Supermassive Black Hole}",
    eprint = "1906.11238",
    archivePrefix = "arXiv",
    primaryClass = "astro-ph.GA",
    doi = "10.3847/2041-8213/ab0ec7",
    journal = "Astrophys. J. Lett.",
    volume = "875",
    pages = "L1",
    year = "2019"
}

@article{Calza2025d,
    author = "Calz{\`a}, Marco and Rinaldi, Massimiliano and Vagnozzi, Sunny",
    title = "{Importance of being nonminimally coupled: Scalar Hawking radiation from regular black holes}",
    eprint = "2510.12257",
    archivePrefix = "arXiv",
    primaryClass = "gr-qc",
    doi = "10.1103/s4cf-bclq",
    journal = "Phys. Rev. D",
    volume = "112",
    number = "10",
    pages = "104055",
    year = "2025"
}

@article{Bhandari2024,
    author = "Bhandari, Gaurav and Pathak, S. D. and Sharma, Manabendra and Khlopov, Maxim Yu",
    title = "{Quantum gravity corrections to Hawking radiation via GUP}",
    eprint = "2407.19268",
    archivePrefix = "arXiv",
    primaryClass = "gr-qc",
    doi = "10.1140/epjp/s13360-025-06608-2",
    journal = "Eur. Phys. J. Plus",
    volume = "140",
    number = "7",
    pages = "663",
    year = "2025"
}

@article{Fathi2021,
    author = "Fathi, Mohsen and Molina, Mart{\'\i}n and Villanueva, J. R.",
    title = "{Adiabatic evolution of Hayward black hole}",
    eprint = "2101.12253",
    archivePrefix = "arXiv",
    primaryClass = "gr-qc",
    doi = "10.1016/j.physletb.2021.136548",
    journal = "Phys. Lett. B",
    volume = "820",
    pages = "136548",
    year = "2021"
}

@article{Calmet:2019eof,
    author = "Calmet, Xavier and Casadio, Roberto and Kuipers, Folkert",
    title = "{Quantum Gravitational Corrections to a Star Metric and the Black Hole Limit}",
    eprint = "1909.13277",
    archivePrefix = "arXiv",
    primaryClass = "hep-th",
    doi = "10.1103/PhysRevD.100.086010",
    journal = "Phys. Rev. D",
    volume = "100",
    number = "8",
    pages = "086010",
    year = "2019"
}

@article{Ellwanger:2021mru,
    author = "Ellwanger, Ulrich",
    title = "{Gradient flow and holography from a local Wilsonian cutoff}",
    eprint = "2110.03371",
    archivePrefix = "arXiv",
    primaryClass = "hep-th",
    doi = "10.1140/epjc/s10052-021-09960-6",
    journal = "Eur. Phys. J. C",
    volume = "81",
    number = "12",
    pages = "1146",
    year = "2021"
}

@article{Odintsov:1991nd,
    author = "Odintsov, S. D. and Shapiro, I. L.",
    title = "{General relativity as the low-energy limit in higher derivative quantum gravity}",
    reportNumber = "PRINT-91-0276 (TOMSK)",
    doi = "10.1088/0264-9381/9/4/006",
    journal = "Class. Quant. Grav.",
    volume = "9",
    pages = "873--882",
    year = "1992"
}

@article{Barvinsky:1994hw,
    author = "Barvinsky, A. O. and Gusev, Yu. V. and Vilkovisky, G. A. and Zhytnikov, V. V.",
    title = "{The Basis of nonlocal curvature invariants in quantum gravity theory. (Third order.)}",
    eprint = "gr-qc/9404061",
    archivePrefix = "arXiv",
    reportNumber = "ALBERTA-THY-14-94",
    doi = "10.1063/1.530427",
    journal = "J. Math. Phys.",
    volume = "35",
    pages = "3525--3542",
    year = "1994"
}

@article{Bonora:2014dfa,
    author = "Bonora, L. and de Brito, K. P. S. and da Rocha, Roldao",
    title = "{Spinor Fields Classification in Arbitrary Dimensions and New Classes of Spinor Fields on 7-Manifolds}",
    eprint = "1411.1590",
    archivePrefix = "arXiv",
    primaryClass = "hep-th",
    reportNumber = "SISSA-57-2014-FISI",
    doi = "10.1007/JHEP02(2015)069",
    journal = "JHEP",
    volume = "02",
    pages = "069",
    year = "2015"
}

@article{Kuntz:2019lzq,
    author = "Kuntz, Iber{\^e} and Casadio, Roberto",
    title = "{Singularity avoidance in quantum gravity}",
    eprint = "1911.05037",
    archivePrefix = "arXiv",
    primaryClass = "hep-th",
    doi = "10.1016/j.physletb.2020.135219",
    journal = "Phys. Lett. B",
    volume = "802",
    pages = "135219",
    year = "2020"
}

@article{Casadio:2021rwj,
    author = "Casadio, Roberto and Kamenshchik, Alexander and Kuntz, Iber{\^e}",
    title = "{Covariant singularities in quantum field theory and quantum gravity}",
    eprint = "2102.10688",
    archivePrefix = "arXiv",
    primaryClass = "hep-th",
    doi = "10.1016/j.nuclphysb.2021.115496",
    journal = "Nucl. Phys. B",
    volume = "971",
    pages = "115496",
    year = "2021"
}

@article{Asorey:2024mkb,
    author = "Asorey, Manuel and Krein, Gastao and Pardina, Miguel and Shapiro, Ilya L.",
    title = "{Bound states of massive complex ghosts in superrenormalizable quantum gravity theories}",
    eprint = "2408.16514",
    archivePrefix = "arXiv",
    primaryClass = "gr-qc",
    doi = "10.1007/JHEP01(2025)113",
    journal = "JHEP",
    volume = "01",
    pages = "113",
    year = "2025"
}

@article{Nastase:2016sji,
    author = "Nastase, Horatiu",
    title = "{Quantum gravity and the holographic dark energy cosmology}",
    eprint = "1602.03708",
    archivePrefix = "arXiv",
    primaryClass = "hep-th",
    doi = "10.1007/JHEP04(2016)149",
    journal = "JHEP",
    volume = "04",
    pages = "149",
    year = "2016"
}

@article{Casadio:2017cdv,
    author = "Casadio, Roberto and Giugno, Andrea and Giusti, Andrea and Lenzi, Michele",
    title = "{Quantum corpuscular corrections to the Newtonian potential}",
    eprint = "1702.05918",
    archivePrefix = "arXiv",
    primaryClass = "gr-qc",
    doi = "10.1103/PhysRevD.96.044010",
    journal = "Phys. Rev. D",
    volume = "96",
    number = "4",
    pages = "044010",
    year = "2017"
}

@article{Casadio:2017sze,
    author = "Casadio, Roberto and Nicolini, Piero and da Rocha, Roldao",
    title = "{Generalised uncertainty principle Hawking fermions from minimally geometric deformed black holes}",
    eprint = "1709.09704",
    archivePrefix = "arXiv",
    primaryClass = "hep-th",
    doi = "10.1088/1361-6382/aad664",
    journal = "Class. Quant. Grav.",
    volume = "35",
    number = "18",
    pages = "185001",
    year = "2018"
}

@article{Donoghue:2018izj,
    author = "Donoghue, John F. and Menezes, Gabriel",
    title = "{Gauge Assisted Quadratic Gravity: A Framework for UV Complete Quantum Gravity}",
    eprint = "1804.04980",
    archivePrefix = "arXiv",
    primaryClass = "hep-th",
    reportNumber = "ACFI- T18-06, ACFI-T18-06, ACFI--T18-06",
    doi = "10.1103/PhysRevD.97.126005",
    journal = "Phys. Rev. D",
    volume = "97",
    number = "12",
    pages = "126005",
    year = "2018"
}

@article{Kuntz:2019omq,
    author = "Kuntz, Iber{\^e} and da Rocha, Rold{\~a}o",
    title = "{One-loop corrections to $\eta/s$ in AdS$_4$/CFT$_3$}",
    eprint = "1909.10121",
    archivePrefix = "arXiv",
    primaryClass = "hep-th",
    doi = "10.1016/j.nuclphysb.2020.115265",
    journal = "Nucl. Phys. B",
    volume = "961",
    pages = "115265",
    year = "2020"
}

@article{Wang:2019rcf,
    author = "Wang, Qingwen and Oshita, Naritaka and Afshordi, Niayesh",
    title = "{Echoes from Quantum Black Holes}",
    eprint = "1905.00446",
    archivePrefix = "arXiv",
    primaryClass = "gr-qc",
    doi = "10.1103/PhysRevD.101.024031",
    journal = "Phys. Rev. D",
    volume = "101",
    number = "2",
    pages = "024031",
    year = "2020"
}

@article{Wang:2025fmz,
    author = "Wang, Zi-Liang and Battista, Emmanuele",
    title = "{Dynamical features and shadows of quantum Schwarzschild black hole in effective field theories of gravity}",
    eprint = "2501.14516",
    archivePrefix = "arXiv",
    primaryClass = "gr-qc",
    doi = "10.1140/epjc/s10052-025-13833-7",
    journal = "Eur. Phys. J. C",
    volume = "85",
    number = "3",
    pages = "304",
    year = "2025"
}

@article{Calmet:2024pve,
    author = "Calmet, Xavier and Casadio, Roberto and Sebastianutti, Marco",
    title = "{Quantum gravitational corrections in cosmology}",
    eprint = "2408.08229",
    archivePrefix = "arXiv",
    primaryClass = "hep-th",
    doi = "10.1016/j.physletb.2024.139208",
    journal = "Phys. Lett. B",
    volume = "860",
    pages = "139208",
    year = "2025"
}

@article{Perrucci:2024qrr,
    author = "Perrucci, Italo and Kuipers, Folkert and Casadio, Roberto",
    title = "{Quantum gravitational hair in gravastars and observational tests}",
    eprint = "2412.04886",
    archivePrefix = "arXiv",
    primaryClass = "gr-qc",
    doi = "10.1088/1475-7516/2025/03/005",
    journal = "JCAP",
    volume = "03",
    pages = "005",
    year = "2025"
}

@article{daRocha:2009gb,
    author = "da Rocha, Roldao and Hoff da Silva, J. M.",
    title = "{ELKO Spinor Fields: Lagrangians for Gravity derived from Supergravity}",
    eprint = "0901.0883",
    archivePrefix = "arXiv",
    primaryClass = "math-ph",
    doi = "10.1142/S0219887809003618",
    journal = "Int. J. Geom. Meth. Mod. Phys.",
    volume = "6",
    pages = "461--477",
    year = "2009"
}

@article{daRocha:2020gee,
    author = "da Rocha, Rold{\~a}o and Tomaz, Anderson A.",
    title = "{MGD-decoupled black holes, anisotropic fluids and holographic entanglement entropy}",
    eprint = "2005.02980",
    archivePrefix = "arXiv",
    primaryClass = "hep-th",
    doi = "10.1140/epjc/s10052-020-8414-8",
    journal = "Eur. Phys. J. C",
    volume = "80",
    number = "9",
    pages = "857",
    year = "2020"
}

@article{daRocha:2021xwq,
    author = "da Rocha, Roldao",
    title = "{Holographic entanglement entropy, deformed black branes, and deconfinement in AdS/QCD}",
    eprint = "2111.01244",
    archivePrefix = "arXiv",
    primaryClass = "hep-th",
    doi = "10.1103/PhysRevD.105.026014",
    journal = "Phys. Rev. D",
    volume = "105",
    number = "2",
    pages = "026014",
    year = "2022"
}

@article{tHooft:1974toh,
    author = "'t Hooft, Gerard and Veltman, M. J. G.",
    title = "{One-loop divergencies in the theory of gravitation}",
    doi = "10.1142/9789814539395_0001",
    journal = "Ann. Inst. H. Poincare Phys. Theor. A",
    volume = "20",
    number = "1",
    pages = "69--94",
    year = "1974"
}

@article{Goroff:1985th,
    author = "Goroff, Marc H. and Sagnotti, Augusto",
    title = "{The Ultraviolet Behavior of Einstein Gravity}",
    reportNumber = "CALT-68-1289, LBL-19995, UCB-PTH-85-34",
    doi = "10.1016/0550-3213(86)90193-8",
    journal = "Nucl. Phys. B",
    volume = "266",
    pages = "709--736",
    year = "1986"
}

@article{Bueno:2016lrh,
    author = "Bueno, Pablo and Cano, Pablo A.",
    title = "{Four-dimensional black holes in Einsteinian cubic gravity}",
    eprint = "1610.08019",
    archivePrefix = "arXiv",
    primaryClass = "hep-th",
    reportNumber = "IFT-UAM-CSIC-16-110",
    doi = "10.1103/PhysRevD.94.124051",
    journal = "Phys. Rev. D",
    volume = "94",
    number = "12",
    pages = "124051",
    year = "2016"
}

@article{Rachwal:2021bgb,
    author = "Rachwal, Leslaw and Modesto, Leonardo and Pinzul, Aleksandr and Shapiro, Ilya L.",
    title = "{Renormalization group in six-derivative quantum gravity}",
    eprint = "2104.13980",
    archivePrefix = "arXiv",
    primaryClass = "hep-th",
    doi = "10.1103/PhysRevD.104.085018",
    journal = "Phys. Rev. D",
    volume = "104",
    number = "8",
    pages = "085018",
    year = "2021"
}

@article{DeFelice:2023vmj,
    author = "De Felice, Antonio and Tsujikawa, Shinji",
    title = "{Excluding static and spherically symmetric black holes in Einsteinian cubic gravity with unsuppressed higher-order curvature terms}",
    eprint = "2305.07217",
    archivePrefix = "arXiv",
    primaryClass = "gr-qc",
    reportNumber = "YITP-23-63, WUCG-23-06",
    doi = "10.1016/j.physletb.2023.138047",
    journal = "Phys. Lett. B",
    volume = "843",
    pages = "138047",
    year = "2023"
}

@article{Lessa:2023thi,
    author = "Lessa, L. A. and Maluf, R. V. and Silva, J. E. G. and Almeida, C. A. S.",
    title = "{Braneworlds in warped Einsteinian cubic gravity}",
    eprint = "2312.06588",
    archivePrefix = "arXiv",
    primaryClass = "gr-qc",
    doi = "10.1088/1475-7516/2024/05/123",
    journal = "JCAP",
    volume = "05",
    pages = "123",
    year = "2024"
}

@article{Matyjasek:2020bzc, 
    author = "Matyjasek, Jerzy",
    title = "{Quasinormal modes of dirty black holes in the effective theory of gravity with a third order curvature term}",
    eprint = "2009.10793",
    archivePrefix = "arXiv",
    primaryClass = "gr-qc",
    doi = "10.1103/PhysRevD.102.124046",
    journal = "Phys. Rev. D",
    volume = "102",
    number = "12",
    pages = "124046",
    year = "2020"
}

@article{Calmet:2023gbw, 
    author = "Calmet, Xavier and Hsu, Stephen D. H. and Sebastianutti, Marco",
    title = "{Quantum gravitational corrections to particle creation by black holes}",
    eprint = "2303.00310",
    archivePrefix = "arXiv",
    primaryClass = "hep-th",
    doi = "10.1016/j.physletb.2023.137820",
    journal = "Phys. Lett. B",
    volume = "841",
    pages = "137820",
    year = "2023"
}

@article{Lambiase:2024vkz, 
    author = {Lambiase, Gaetano and Pantig, Reggie C. and {\"O}vg{\"u}n, Ali},
    title = "{Traces of quantum gravitational correction at third-order curvature through the black hole shadow and particle deflection at the weak field limit}",
    eprint = "2406.18292",
    archivePrefix = "arXiv",
    primaryClass = "gr-qc",
    doi = "10.1016/j.dark.2024.101597",
    journal = "Phys. Dark Univ.",
    volume = "46",
    pages = "101597",
    year = "2024"
}

@article{Marciu:2023hdb, 
    author = "Marciu, Mihai and Ioan, Dana Maria and Dragomir, Mihai",
    title = "{Observational constraints for cubic gravity theory based on third order contractions of the Riemann tensor}",
    eprint = "2311.11297",
    archivePrefix = "arXiv",
    primaryClass = "gr-qc",
    doi = "10.1140/epjc/s10052-024-12559-2",
    journal = "Eur. Phys. J. C",
    volume = "84",
    number = "2",
    pages = "196",
    year = "2024"
}

@article{Wang2025a,
    author = "Wang, Zi-Liang",
    title = "{Exploring the role of accretion disk geometry in shaping black hole shadows}",
    eprint = "2506.21148",
    archivePrefix = "arXiv",
    primaryClass = "gr-qc",
    doi = "10.1103/fhqj-wgcm",
    journal = "Phys. Rev. D",
    volume = "112",
    number = "6",
    pages = "064052",
    year = "2025"
}

@article{Casadio:2021gdf,
    author = "Casadio, Roberto and Giusti, Andrea and Kuntz, Iber{\^e} and Neri, Giulio",
    title = "{Effective metric outside bootstrapped Newtonian sources}",
    eprint = "2101.12471",
    archivePrefix = "arXiv",
    primaryClass = "gr-qc",
    doi = "10.1103/PhysRevD.103.064001",
    journal = "Phys. Rev. D",
    volume = "103",
    pages = "064001",
    year = "2021"
}

@article{Muck:2016stv,
    author = {M{\"u}ck, Wolfgang},
    title = "{Hawking radiation is corpuscular}",
    eprint = "1606.01790",
    archivePrefix = "arXiv",
    primaryClass = "hep-th",
    doi = "10.1140/epjc/s10052-016-4233-3",
    journal = "Eur. Phys. J. C",
    volume = "76",
    number = "7",
    pages = "374",
    year = "2016"
}

@article{Casadio:2021eio,
    author = "Casadio, Roberto",
    title = "{Geometry and thermodynamics of coherent quantum black holes}",
    eprint = "2103.00183",
    archivePrefix = "arXiv",
    primaryClass = "gr-qc",
    doi = "10.1142/S0218271822501280",
    journal = "Int. J. Mod. Phys. D",
    volume = "31",
    number = "16",
    pages = "2250128",
    year = "2022"
}

@article{Casadio:2015bna,
    author = "Casadio, Roberto and Giugno, Andrea and Orlandi, Alessio",
    title = "{Thermal corpuscular black holes}",
    eprint = "1504.05356",
    archivePrefix = "arXiv",
    primaryClass = "gr-qc",
    doi = "10.1103/PhysRevD.91.124069",
    journal = "Phys. Rev. D",
    volume = "91",
    number = "12",
    pages = "124069",
    year = "2015"
}

@article{Giusti:2019wdx,
    author = "Giusti, Andrea",
    title = "{On the corpuscular theory of gravity}",
    doi = "10.1142/S0219887819300010",
    journal = "Int. J. Geom. Meth. Mod. Phys.",
    volume = "16",
    number = "03",
    pages = "1930001",
    year = "2019"
}

@article{Casadio:2016zpl,
    author = "Casadio, Roberto and Giugno, Andrea and Giusti, Andrea",
    title = "{Matter and gravitons in the gravitational collapse}",
    eprint = "1606.04744",
    archivePrefix = "arXiv",
    primaryClass = "hep-th",
    doi = "10.1016/j.physletb.2016.10.058",
    journal = "Phys. Lett. B",
    volume = "763",
    pages = "337--340",
    year = "2016"
}

@article{Dvali:2011aa,
    author = "Dvali, Gia and Gomez, Cesar",
    title = "{Black Hole's Quantum N-Portrait}",
    eprint = "1112.3359",
    archivePrefix = "arXiv",
    primaryClass = "hep-th",
    doi = "10.1002/prop.201300001",
    journal = "Fortsch. Phys.",
    volume = "61",
    pages = "742--767",
    year = "2013"
}

@article{Dvali:2012rt,
    author = "Dvali, Gia and Gomez, Cesar",
    title = "{Black Hole's 1/N Hair}",
    eprint = "1203.6575",
    archivePrefix = "arXiv",
    primaryClass = "hep-th",
    doi = "10.1016/j.physletb.2013.01.020",
    journal = "Phys. Lett. B",
    volume = "719",
    pages = "419--423",
    year = "2013"
}

@article{Casadio:2015jha,
    author = "Casadio, Roberto and Cavalcanti, Rogerio T. and Giugno, Andrea and Mureika, Jonas",
    title = "{Horizon of quantum black holes in various dimensions}",
    eprint = "1509.09317",
    archivePrefix = "arXiv",
    primaryClass = "gr-qc",
    doi = "10.1016/j.physletb.2016.06.042",
    journal = "Phys. Lett. B",
    volume = "760",
    pages = "36--44",
    year = "2016"
}

@article{Casadio:2022ndh,
    author = "Casadio, Roberto and Giusti, Andrea and Ovalle, Jorge",
    title = {{Quantum Reissner-Nordstr{\"o}m geometry: Singularity and Cauchy horizon}},
    eprint = "2203.03252",
    archivePrefix = "arXiv",
    primaryClass = "gr-qc",
    doi = "10.1103/PhysRevD.105.124026",
    journal = "Phys. Rev. D",
    volume = "105",
    number = "12",
    pages = "124026",
    year = "2022"
}

@article{Cadoni:2018dnd,
    author = "Cadoni, M. and Casadio, R. and Giusti, A. and Tuveri, M.",
    title = "{Emergence of a Dark Force in Corpuscular Gravity}",
    eprint = "1801.10374",
    archivePrefix = "arXiv",
    primaryClass = "gr-qc",
    doi = "10.1103/PhysRevD.97.044047",
    journal = "Phys. Rev. D",
    volume = "97",
    number = "4",
    pages = "044047",
    year = "2018"
}

@article{Casadio:2024whh,
    author = "Casadio, R.",
    title = "{Quantum Rotating Black Holes (Recovering Geometry in a Quantum World)}",
    doi = "10.15407/ujpe69.7.466",
    journal = "Ukr. J. Phys.",
    volume = "69",
    number = "7",
    pages = "466",
    year = "2024"
}

@article{Feng:2024bsx,
    author = "Feng, Wenbin and da Rocha, Roldao and Casadio, Roberto",
    title = "{Hair and entropy for slowly rotating quantum black holes}",
    eprint = "2401.14540",
    archivePrefix = "arXiv",
    primaryClass = "gr-qc",
    doi = "10.1140/epjc/s10052-024-12817-3",
    journal = "Eur. Phys. J. C",
    volume = "84",
    number = "6",
    pages = "586",
    year = "2024"
}

@article{Calmet:2023met,
    author = "Calmet, Xavier and Casadio, Roberto and Hsu, Stephen D. H. and Kuipers, Folkert",
    title = "{Quantum hair during gravitational collapse}",
    eprint = "2305.09466",
    archivePrefix = "arXiv",
    primaryClass = "hep-th",
    doi = "10.1103/PhysRevD.108.086012",
    journal = "Phys. Rev. D",
    volume = "108",
    number = "8",
    pages = "086012",
    year = "2023"
}

@article{Casadio:2024lzd,
    author = "Casadio, R. and da Rocha, R. and Giusti, A. and Meert, P.",
    title = "{Black holes with a charged quantum dust core}",
    eprint = "2407.04146",
    archivePrefix = "arXiv",
    primaryClass = "gr-qc",
    doi = "10.1103/PhysRevD.110.104067",
    journal = "Phys. Rev. D",
    volume = "110",
    number = "10",
    pages = "104067",
    year = "2024"
}

@article{Casadio:2023pmh,
    author = "Casadio, R. and da Rocha, R. and Giusti, A. and Meert, P.",
    title = "{Thermodynamic and configurational entropy of quantum Schwarzschild geometries}",
    eprint = "2310.07505",
    archivePrefix = "arXiv",
    primaryClass = "gr-qc",
    doi = "10.1016/j.physletb.2024.138466",
    journal = "Phys. Lett. B",
    volume = "849",
    pages = "138466",
    year = "2024"
}

@article{Meert:2025cct,
    author = "Meert, Pedro and Giusti, Andrea and Casadio, Roberto",
    title = "{Quantum maximally symmetric spacetimes}",
    eprint = "2504.02786",
    archivePrefix = "arXiv",
    primaryClass = "gr-qc",
    doi = "10.1016/j.physletb.2025.139613",
    journal = "Phys. Lett. B",
    volume = "867",
    pages = "139613",
    year = "2025"
}

@article{Antonelli:2025mcv,
    author = "Antonelli, Tommaso and Sebastianutti, Marco and Giusti, Andrea",
    title = "{Coherent electrically-charged quantum black holes}",
    eprint = "2506.02231",
    archivePrefix = "arXiv",
    primaryClass = "gr-qc",
    doi = "10.1140/epjc/s10052-025-14977-2",
    journal = "Eur. Phys. J. C",
    volume = "85",
    number = "10",
    pages = "1219",
    year = "2025"
}

@article{Feng:2025nai,
    author = "Feng, Wenbin and Giusti, Andrea and Casadio, Roberto",
    title = "{Horizon quantum mechanics for coherent quantum black holes}",
    eprint = "2408.17091",
    archivePrefix = "arXiv",
    primaryClass = "gr-qc",
    doi = "10.1140/epjp/s13360-025-06065-x",
    journal = "Eur. Phys. J. Plus",
    volume = "140",
    number = "2",
    pages = "145",
    year = "2025"
}

@article{Urmanov:2024qai,
    author = "Urmanov, Abdybek and Chakrabarty, Hrishikesh and Malafarina, Daniele",
    title = "{Observational properties of coherent quantum black holes}",
    eprint = "2406.04813",
    archivePrefix = "arXiv",
    primaryClass = "gr-qc",
    doi = "10.1103/PhysRevD.110.044030",
    journal = "Phys. Rev. D",
    volume = "110",
    number = "4",
    pages = "044030",
    year = "2024"
}

@article{Cirilo-Lombardo:2023cxb,
    author = "Cirilo-Lombardo, Diego J. and Sanchez, Norma G.",
    title = "{Coherent states of quantum spacetimes for black holes and de Sitter spacetime}",
    eprint = "2312.06628",
    archivePrefix = "arXiv",
    primaryClass = "gr-qc",
    doi = "10.1103/PhysRevD.108.126001",
    journal = "Phys. Rev. D",
    volume = "108",
    number = "12",
    pages = "126001",
    year = "2023"
}

@book{Faraoni:2015ula,
    author = "Faraoni, Valerio",
    title = "{Cosmological and Black Hole Apparent Horizons}",
    doi = "10.1007/978-3-319-19240-6",
    isbn = "978-3-319-19239-0, 978-3-319-19240-6",
    publisher={Springer},
    year = "2015"
}

@article{Mazur:2004fk,
    author = "Mazur, Pawel O. and Mottola, Emil",
    title = "{Gravitational vacuum condensate stars}",
    eprint = "gr-qc/0407075",
    archivePrefix = "arXiv",
    doi = "10.1073/pnas.0402717101",
    journal = "Proc. Nat. Acad. Sci.",
    volume = "101",
    pages = "9545--9550",
    year = "2004"
}

@article{Jumaniyozov:2025,
    author = "Jumaniyozov, Shokhzod and Zahid, Muhammad and Alloqulov, Mirzabek and Ibragimov, Inomjon and Rayimbaev, Javlon and Murodov, Sardor",
    title = "{Radiative properties and QPOs around charged black hole in Kalb{\textendash}Ramond gravity}",
    doi = "10.1140/epjc/s10052-025-13863-1",
    journal = "Eur. Phys. J. C",
    volume = "85",
    number = "2",
    pages = "126",
    year = "2025"
}

@article{Murodov:2026,
    author = "Murodov, Sardor and Rahmatov, Bekzod and Umarov, Otabek and Kurbaniyazov, Anvar and Javed, Faisal and Rayimbaev, Javlon",
    title = "{Circular-orbit dynamics and QPO constraints in static Einstein-scalar-Gauss-Bonnet black holes}",
    eprint = "2607.19805",
    archivePrefix = "arXiv",
    primaryClass = "gr-qc",
    doi = "10.1140/epjc/s10052-026-16256-0",
    journal = "Eur. Phys. J. C",
    volume = "86",
    number = "8",
    pages = "1007",
    year = "2026"
}

@article{Turakhonov:2026,
    author = "Turakhonov, Ziyodulla and Hoshimov, Husanboy and Davlataliev, Akbar and Atamurotov, Farruh and Abdujabbarov, Ahmadjon and Yuan, Chengxun",
    title = "{Particle dynamics and quasi-periodic oscillations in the non-minimally coupled Horndeski black hole}",
    doi = "10.1016/j.nuclphysb.2026.117566",
    journal = "Nucl. Phys. B",
    volume = "1029",
    pages = "117566",
    year = "2026"
}

@article{Zhang:2026snap,
    author = "Zhang, J. and Huang, W. and Wang, Q. and Shaw, A. D. and Friswell, M. I. and Huang, K.",
    title = "{Exploring sequential snapping bifurcation through a tunable energy landscape}",
    doi = "10.1103/nxpg-92kv",
    journal = "Phys. Rev. Applied",
    year = "2026"
}

@article{Xiang:2026blazar,
    author = "Xiang, Y. and Ye, Y. and Feng, P. and Li, H. and Pang, X. and Lan, X. and Zhang, N.",
    title = "{A novel pipeline for the identification of new gamma-ray blazars from the 4FGL--Xiang catalog based on multiwavelength flux distributions}",
    doi = "10.1093/mnras/stag150",
    journal = "Mon. Not. Roy. Astron. Soc.",
    volume = "546",
    number = "4",
    pages = "stag150",
    year = "2026"
}

@article{Zhang:2026doppler,
    author = "Zhang, Y. and Ba, D. and Yang, Y. and Dong, Y.",
    title = "{Generalized Doppler effect for high-accuracy frequency shift measurement}",
    doi = "10.1038/s41377-026-02259-9",
    journal = "Light Sci. Appl.",
    volume = "15",
    number = "1",
    pages = "197",
    year = "2026"
}

@article{Septembre:2025drift,
    author = "Septembre, I. and Liu, Z. and Kokhanchik, P. and Liang, S. and Liu, F. and Li, C. and Zhang, Z.",
    title = "{Transverse Non-Hermitian Drift Induced by the Quantum Metric of Exceptional Rings}",
    doi = "10.1103/m978-tbzc",
    journal = "Phys. Rev. Lett.",
    volume = "135",
    number = "26",
    pages = "263802",
    year = "2025"
}

@article{Li:2025nonreciprocal,
    author = "Li, C. and Yuan, J. and He, R. and Yu, J. and Zhang, Y. and Xiao, M. and Zhang, Z.",
    title = "{Nonreciprocal spontaneous parametric process}",
    doi = "10.1038/s41377-025-01844-8",
    journal = "Light Sci. Appl.",
    volume = "14",
    number = "1",
    pages = "200",
    year = "2025"
}

@article{Liang:2025localization,
    author = "Liang, S. and Li, C. and Liu, Z. and Yu, W. and Li, C. and Zhang, Y. and Zhang, Z.",
    title = "{Observation of nonreciprocal transverse localization of light}",
    doi = "10.1038/s41467-025-65084-5",
    journal = "Nature Commun.",
    volume = "16",
    number = "1",
    pages = "10248",
    year = "2025"
}

@article{Li:2022eue,
    author = "Li, Song and Mirzaev, Temurbek and Abdujabbarov, Ahmadjon A. and Malafarina, Daniele and Ahmedov, Bobomurat and Han, Wen-Biao",
    title = "{Constraining the deformation of a rotating black hole mimicker from its shadow}",
    eprint = "2207.10933",
    archivePrefix = "arXiv",
    primaryClass = "gr-qc",
    doi = "10.1103/PhysRevD.106.084041",
    journal = "Phys. Rev. D",
    volume = "106",
    number = "8",
    pages = "084041",
    year = "2022"
}

@article{Mirzaev:2023oud,
    author = "Mirzaev, Temurbek and Li, Song and Narzilloev, Bakhtiyor and Hussain, Ibrar and Abdujabbarov, Ahmadjon and Ahmedov, Bobomurat",
    title = "{Simulated image of the shadow of the Kerr{\textendash}Newman{\textendash}NUT{\textendash}Kiselev black hole in the Rastall gravity with a thin accretion disk}",
    doi = "10.1140/epjp/s13360-022-03632-4",
    journal = "Eur. Phys. J. Plus",
    volume = "138",
    number = "1",
    pages = "47",
    year = "2023"
}

@article{Mirzaev:2024qcu,
    author = "Mirzaev, Temurbek and Bambi, Cosimo and Abdikamalov, Askar B. and Jiang, Jiachen and Liu, Honghui and Riaz, Shafqat and Shashank, Swarnim",
    title = "{X-Ray Spectra of Black Hole X-Ray Binaries with Returning Radiation}",
    eprint = "2406.01226",
    archivePrefix = "arXiv",
    primaryClass = "astro-ph.HE",
    doi = "10.3847/1538-4357/ad8a63",
    journal = "Astrophys. J.",
    volume = "976",
    number = "2",
    pages = "229",
    year = "2024"
}

@article{Mukazhanov:2024rka,
    author = "Mukazhanov, Olzhas and Roy, Rittick and Mirzaev, Temurbek and Bambi, Cosimo",
    title = "{Numerical parametrization of stationary axisymmetric black holes in a theory agnostic framework}",
    eprint = "2404.17055",
    archivePrefix = "arXiv",
    primaryClass = "gr-qc",
    doi = "10.1103/PhysRevD.110.024060",
    journal = "Phys. Rev. D",
    volume = "110",
    number = "2",
    pages = "024060",
    year = "2024"
}

@article{Shashank:2024hsi,
    author = "Shashank, Swarnim and Abdikamalov, Askar B. and Liu, Honghui and Mirzaev, Temurbek and Jiang, Jiachen and Bambi, Cosimo and Baker, Fergus and Young, Andrew",
    title = "{Modeling Reflection Spectra of Super-Eddington X-Ray Sources}",
    eprint = "2407.12890",
    archivePrefix = "arXiv",
    primaryClass = "astro-ph.HE",
    doi = "10.3847/1538-4357/ae2614",
    journal = "Astrophys. J.",
    volume = "996",
    number = "1",
    pages = "51",
    year = "2026"
}

@article{Errehymy:2026bbl,
    author = "Errehymy, A. and Khedif, Y. and Daoud, M. and Turimov, B. and Usanov, S. and Turaev, F. and Yasakov, Z.",
    title = "{Joint constraints on quantum Oppenheimer{\textendash}Snyder black holes from high-frequency quasi-periodic oscillations in X-ray binary systems and the S2 star orbit}",
    doi = "10.1016/j.physletb.2026.140930",
    journal = "Phys. Lett. B",
    volume = "881",
    pages = "140930",
    year = "2026"
}

@article{Errehymy:2026wti,
    author = "Errehymy, A. and Hansraj, S. and Hansraj, C.",
    title = "{Observational Limits on Einasto Dark Matter Parameters from Event Horizon Telescope Images of Sgr A$^{∗}$ and M87$^{∗}$}",
    eprint = "2607.07752",
    archivePrefix = "arXiv",
    primaryClass = "gr-qc",
    doi = "10.3847/1538-4357/ae88f0",
    journal = "Astrophys. J.",
    volume = "1008",
    number = "1",
    pages = "78",
    year = "2026"
}

@article{Errehymy:2026fgm,
    author = "Errehymy, A. and Khedif, Y. and Daoud, M. and Turimov, B. and Usanov, S. and Yasakov, Z. and Avezmuratova, Z.",
    title = "{Observational constraints on Kazakov-Solodukhin quantum deformed black holes from M87* and Sgr A* shadows}",
    eprint = "2607.02631",
    archivePrefix = "arXiv",
    primaryClass = "gr-qc",
    doi = "10.1016/j.nuclphysb.2026.117579",
    journal = "Nucl. Phys. B",
    volume = "1029",
    pages = "117579",
    year = "2026"
}

\end{document}